\documentclass{article}

\usepackage{arxiv}

\usepackage[utf8]{inputenc} 
\usepackage[T1]{fontenc}    
\usepackage{hyperref}       
\usepackage{url}            
\usepackage{booktabs}       
\usepackage{amsfonts}       
\usepackage{nicefrac}       
\usepackage{microtype}      
\usepackage{lipsum}
\usepackage{graphicx}
\usepackage{fancyhdr}
\usepackage[most]{tcolorbox}
\usepackage{pgfplots}
\usepackage{longtable}
\graphicspath{ {./images/} }

\title{The current state of security --- \\ Insights from the German software industry}

\author{
  Timo Langstrof \\
  Research Group Software Construction\\
  RWTH Aachen University\\
  Ahornstraße 55 \\
  Aachen, Germany \\
  \texttt{timo.langstrof@rwth-aachen.de}
  \And
  Alex R. Sabau \\
  Research Group Software Construction\\
  RWTH Aachen University\\
  Ahornstraße 55 \\
  Aachen, Germany \\
  \texttt{sabau@swc.rwth-aachen.de} \\
}

\begin{document}
\maketitle
\thispagestyle{fancy}
\begin{abstract}
These days, software development and security go hand in hand. Numerous techniques and strategies are discussed in the literature that can be applied to guarantee the incorporation of security into the software development process. In this paper, the main ideas of secure software development that have been discussed in the literature are outlined. Next, a dataset on implementation in practice is gathered through a qualitative interview study involving 20 companies. Trends and correlations in this dataset are found and contrasted with theoretical ideas from the literature. The results show that the organizations that were polled are placing an increasing focus on security. Although the techniques covered in the literature are used in the real world, they are frequently not fully integrated into formal, standardized processes. The insights gained from our research lay the groundwork for future research, which can delve deeper into specific elements of these methods to enhance our understanding of their application in real-world scenarios. \footnote{This paper is largely an extract from a bachelor thesis and was extended by some parts as a preparation for a journal publication. We decided to publish a preprint of the paper to provide access to our data early on. Interview transcripts have been removed for reasons of anonymization. An adapted version of the transcripts will be added to the journal publication.}
\end{abstract}
\newtcolorbox{mainBox}[1]{%
		enhanced,
		colback=white!90!black,
		colframe=darkgray,
		colbacktitle=darkgray,
		coltitle=white,
		fonttitle=\bfseries,
		title=#1,
		rounded corners,
		boxrule=3pt,
		drop shadow=darkgray,
		width=\textwidth,
	}

\section{Introduction} \label{chap:Introduction}
\noindent In recent years, digitization has assumed a significant role in both the industry and daily life and continues to evolve steadily. Software products have become an integral part of many areas of life, from the automotive industry and healthcare to the finance sector \cite{Sim05}. Parallel to this, the subject of cybersecurity is gaining steadily in importance, a trend amplified by the increasing demand for software and the rapid advancement of technologies  \cite{Broy.2006} \cite{cenk23}. These generate a heightened level of complexity in software systems \cite{dyb09}, continually opening up new possibilities and targets for attacks \cite{cenk23}.

The urgency of secure software development is underscored by the increasing number of cyber-attacks and security breaches \cite{olga22}. These not only lead to substantial financial and reputational damage to the affected companies but also can significantly impact their end-users. Whether through direct financial losses, the loss or release of personal data, or the interruption of critical software processes, the consequences are far-reaching and severe \cite{mamdouh20}.

Therefore, the development of secure software is experiencing an increasing presence in industry and research \cite{bsimm}. As a means to improve the security of software systems, the ``security by design'' paradigm emerged in recent years, which is a promising approach to security as it promotes to implement security from the very start of the software development life cycle, rather than later on \cite{Secure_by_design2019, TOGAF9.1_Security_and_ADM}. It is essential that these developments are also reflected in the practices and strategies of the companies that develop and operate software. Security in software development can no longer be considered secondary. It must be understood as a central component of software development, equivalent to all other functionalities \cite{SALINI20121785}.

It is crucial that companies provide not only the necessary resources and knowledge for secure software development but also create an awareness of the importance of cybersecurity in all phases of software development \cite{9669954}. This also includes the continuous education and sensitization of employees and the implementation of effective and sustainable security policies and processes \cite{bsimm}.

\subsection{Research questions} \label{sec:ResearchQuestions}
In this study, the focus is on the identification and analysis of current processes and methods employed by companies to ensure the security of developed software. The objective is to gain profound insights into potential trends, patterns, and, most importantly, discrepancies between theoretical concepts and actually implemented practices in software development. The theoretical perspective and its relevance are critically examined in the context of the practical implementation of security concepts in software development to identify existing gaps between theory and practice. From this objective, the central research question arises.\\ \\
\textbf{(RQ1):} \label{rq1} \textit{To what extent do the models and methods for secure software development presented in the literature differ from the practices actually applied in the industry?}\\

To answer this central question in a detailed and comprehensive manner, the subsequent sub-question is initially addressed. This aims to develop a clearly defined understanding of which processes and methods are considered standard for secure software development by the scientific community. The answers to this question form the basis for precisely analyzing the relationship between theoretical specifications and practical implementation.\\ \\
\textbf{(RQ2):} \label{rq2} \textit{Which theoretical models and methods for secure software development are represented in the current scientific literature?}\\

Similarly, to deepen the analysis of the central research question, the third sub-question is of importance. It allows for practical insights into the industrial processes and strategies for securing software, thereby enabling a comprehensive assessment of the applicability and effectiveness of theoretical models in practice.\\ \\
\textbf{(RQ3):} \label{rq3} \textit{What  processes and measures do companies in the industry implement  to ensure the development of secure software, and what strategies do they pursue to ensure the ongoing security of software products?}

Finally, as security by design is often promoted to positively affect the security of software systems, our interviews shall identify trends in the extend to which the this paradigm is being pursued in industry. \\ \\
\textbf{(RQ4):} \label{rq4} \textit{To what extend is the ``security by design'' paradigm being followed in  the software industry?}

\subsection{Structure of this paper} \label{sec:Structure}

This paper primarily aims to investigate how security in software development is implemented in practice by companies. To this end, it is divided into three parts. The first part establishes a theoretical foundation on the topic to gain insight into the methods considered in the literature for secure software development. The second part involves designing and implementing a survey in the industry to create a data basis for the investigation. In the third and final part, these data are evaluated and compared with the findings of the theoretical foundation.

The first part includes Chapters \ref{chap:RelatedWork} and \ref{chap:Theoretical}. Chapter \ref{chap:RelatedWork} examines the existing research in this area and what can be built upon. This involves identifying existing literature reviews and studies with companies. Chapter \ref{chap:Theoretical} builds on the results of these studies. It addresses the second research question RQ2 by identifying and describing the most important aspects of secure software development from existing studies. This creates a foundation for answering the central research question RQ1 later on. Chapter \ref{chap:Methology} deals with the planning and execution of the study in the industry. It examines the approach used to collect data, the data points to be collected, and how the study was conducted. Chapter \ref{chap:Analysis} then evaluates and presents the data collected. This involves three steps: presentation, interpretation, and relation to the literature. The first two of these steps answer the research questions RQ3 and RQ4, which, together with the results of research question RQ2, leads to the answer to research question RQ1 in the third step. To conclude, Chapter \ref{chap:Conclusion} summarizes the results of the work, and a brief outlook on possible future works is provided, encapsulating the insights gained into the alignment or discrepancies between theory and practice in the field of secure software development.

\section{Related Work} \label{chap:RelatedWork}
\noindent In this chapter, we examine other studies that have been conducted on this topic. This includes research focused on secure software development as discussed in literature, as well as studies involving surveys with companies.

\subsection{Survey and analysis on Security Requirements Engineering}
The 2012 study `Survey and analysis on Security Requirements Engineering' \cite{SALINI20121785} by P. Salini and S. Kanmani focuses on how security requirements and their processes are considered in the literature. They first identify the different types of security requirements, namely identification, authentication, authorization, immunity, integrity, intrusion detection, non-repudiation, privacy, security auditing, survivability, physical protection, and system maintenance security, and provide descriptions and examples of possible requirements. They conclude that security requirements should no longer be seen as non-functional requirements and that it is of high importance to raise them as early as possible in the Software Development Life Cycle (SDLC).

Subsequently, they describe the steps necessary for gathering these requirements. The central points they identify are Threat Modeling, Risk Analysis, and Requirements Review.

In the second part of the study, Salini and Kanmani examine the processes for gathering security requirements existing in the literature and compare their methodologies. The processes they look at include Security Quality Requirements Engineering (SQUARE) \cite{square}, the framework by Haley and his colleagues \cite{haley2006}, Gustav Boström's Security Requirements Engineering in extreme programming \cite{bostroem2006}, Comprehensive, Lightweight Application Security Process (CLASP) \cite{graham2006introduction}, Microsoft Trustworthy Computing Security Development Life Cycle (MTCSDLC) \cite{MTCSDLC}, Security Requirements Engineering Process (SREP) \cite{MELLADO2007244}, and Secure Tropos \cite{Mouratidis.2007} . They examine these processes to see which relevant methodologies for gathering security requirements they follow and at which phases of the SDLC they can be applied. They conclude that SQUARE and SREP offer the best options for companies, as they encompass all the steps.
\subsection{Systematic Literature Review on Security Risks and Its Practices in Secure Software Development}
A. Khan and colleagues carried out a systematic literature review on security in the SDLC in 2022 under the title "Systematic Literature Review on Security Risks and Its Practices in Secure Software Development" \cite{9669954}. Their study's objectives are to pinpoint security risks and threats in the SDLC and to emphasize the security best practices that must be included into each stage of the SDLC in order to increase the security of the development process.

Six digital libraries were searched using a search query to find pertinent scientific papers for the SLR. ACM Digital Library, IEEE-Xplore, Sciencedirect, Springer Link, Wiley Online Library, and Google Scholar were among them. Papers published in peer-reviewed publications and conferences between 2000 and 2020 were used. After applying exclusion criteria, 121 of the 12114 papers that were found were used for evaluation.

The subsequent stage involved determining which risks and practices were highlighted, as well as which stages of the SDLC (Requirements Engineering, Design, Coding, Testing, Deployment, and Management) were most commonly cited in the papers.  According to A. Khan et al., the requirements engineering stage takes into account a sizable percentage (97.5\%) of the security threats that have been discovered. They come to the conclusion that a lot of software security problems are caused by inadequate or inaccurate security requirement identification, documentation, analysis, mapping, priority, specification, and availability.
\subsection{BSIMM}
The Building Security In Maturity Model (BSIMM) \cite{bsimm} was first published in 2008 by Dr. Gary McGraw, Dr. Sammy Migues, and Dr. Brian Chess. Since then, this model has been regularly revised and published. The most current version at the time of our research is BSIMM13, which was released in 2022. The model is a framework for companies to examine and assess their security process in order to find and fix potential vulnerabilities in it. The model was developed by the authors creating a framework, the software security framework. They then conduct interviews with people working in secure software development. This identifies general activities and approaches in secure software development.

In BSIMM13, 125 activities were identified through interviews with 130 companies. These activities are divided into areas. The authors describe the areas as follows:
\begin{itemize}
	\item \textbf{Governance}: Practices that help organize, manage, and measure a software security initiative. Staff development is also a central governance practice.
	\item \textbf{Intelligence}: Practices that result in collections of corporate knowledge used in carrying out software security activities throughout the organization. Collections include both proactive security guidance and organizational threat modeling.
	\item \textbf{SSDL Touchpoints}: Practices associated with analysis and assurance of particular software development artifacts and processes. All software security methodologies include these practices.
	\item \textbf{Deployment}: Practices that interface with traditional network security and software maintenance organizations. Software configuration, maintenance, and other environment issues have a direct impact on software security.
\end{itemize}
When a company wants to use this framework to examine its own approach, it determines which of the activities it carries out. Together with the frequency of the activity in the surveyed companies, a score is then calculated for each area. With this, companies can identify where they need to improve and in which areas they are on a good footing.

\section{Literature} \label{chap:Theoretical}
\noindent
This section aims to identify the fundamental approaches and standards for secure software development from the literature. It explores three different areas relevant for the study based on the Software Development Life Cycle \cite{SALINI20121785}:
\begin{itemize}
	\item Defining security requirements
	\item Handling of security vulnerabilities
	\item Common standards and guidelines
\end{itemize}
As already observed in Chapter \ref{chap:RelatedWork}, there are already numerous SLRs on this topic. Therefore, another SLR would only make a limited scientific contribution. Furthermore, the primary goal of our research is the execution and evaluation of the Qualitative Study with the companies. Therefore, we have decided to build the investigation of the theoretical foundations on the basis of existing SLRs.

\subsection{Security Requirements} \label{theo_sr}
Security requirements describe the functions and constraints that software must fulfill in terms of security. These cover various areas of security, such as authentication or identification. It should be noted that security requirements should be considered as functional requirements \cite{SALINI20121785}. To capture these requirements, a security investigation of the software should be carried out. As this is a critical component of software development, this should happen as early as possible in the development process \cite{MELLADO2007244} \cite{SALINI20121785}.
The study 'Survey and analysis on Security Requirements Engineering' by P. Salini and S. Kanmani \cite{SALINI20121785} identifies some possible approaches for collecting these requirements. These methods include the SQUARE process \cite{square}, CLASP \cite{DEWIN20091152} \cite{graham2006introduction}, Microsoft Trustworthy Computing Security Development Life Cycle (MTCSDLC) \cite{MTCSDLC}, and SREP \cite{MELLADO2007244}. These approaches all describe a similar process but focus on different aspects. For example, the MTCSD focuses on the integration of customer requirements \cite{MTCSDLC}. CLASP is developed by the Open Web Application Security Project (OWASP) community and therefore focuses particularly on security in web applications, relying on a comprehensive catalog of security best practices as the basis for the requirements \cite{DEWIN20091152}.

Across all these approaches, three core activities can be identified that should be fulfilled in creating security requirements.

\subsubsection{Threat Modeling} \label{theo_tm}
Threat Modeling means that developers model all possible threats to the software by viewing it from the perspective of a potential attacker to determine what could go wrong  \cite{SALINI20121785} \cite{XIONG201953} The most common approach here is the STRIDE method \cite{ELHADARY2014463} developed by Microsoft \cite{SALINI20121785}. The name stands for six different types of security threats: Spoofing, Tampering, Repudiation, Information Disclosure, Denial of Service, and Elevation of Privilege.
\subsubsection{Risk Analysis} \label{theo_ra}
In risk analysis, the vulnerabilities found through threat modeling are assessed. This is based on the fact that not all vulnerabilities are equally critical and should therefore be prioritized differently \cite{1492335}. There are various approaches to assessing the risk. A common method is the Dread classification. Each of the identified vulnerabilities is assessed in relation to these five aspects  \cite{suprihanto2018determination}:
\begin{itemize}
	\item \textbf{Damage}: Assesses the potential extent of the damage that successful exploitation of the vulnerability could cause. This includes considering factors such as data loss, system interruption, or reputational damage. The assessment is often made on a scale, for example, from 0 (no damage) to 10 (catastrophic damage).
	\item \textbf{Reproducibility}: Measures how easily the vulnerability can be reproduced. A high rating is given if the vulnerability can be easily and consistently exploited by an attacker without special conditions or advanced knowledge.
	\item \textbf{Exploitability}: Assesses how easily the vulnerability can be exploited. This includes considering the required tools, skills, and resources of a potential attacker. A high rating indicates that the vulnerability is easy to exploit.
	\item \textbf{Affected Users}: Estimates the percentage of users or systems that would be affected by the vulnerability if it were exploited. A higher rating is given if a larger proportion of the user base is affected.
	\item \textbf{Discoverability}: Refers to how easily an attacker can discover the vulnerability. This can be influenced by the availability of information about the vulnerability or the ease with which an attacker can identify the vulnerability.
\end{itemize}
These values are then calculated together to obtain an overall assessment of the vulnerability \cite{suprihanto2018determination}.

\subsubsection{Requirement Review} \label{theo_srep}
In this process, an external catalog of security requirements is used as the basis for analysis. The chosen catalog is filtered for the specific application area. This approach ensures that a wide range of requirements is covered while simultaneously not having to redefine the requirements for each project, thereby increasing efficiency \cite{MELLADO2007244} \cite{SALINI20121785}.

\subsubsection{Security Vulnerabilities} \label{theo_sv}
A key aspect in the development of secure software is the handling of emerging security vulnerabilities. In the study of Williams, McGraw, and Migues "Engineering Security Vulnerability Prevention, Detection, and Response" \cite{8409917} two critical aspects have been identified that are essential in managing these vulnerabilities: Detection and Response.
\subsubsection{Detection of Security Vulnerabilities}
The detection of security vulnerabilities is a crucial step in identifying weaknesses within the code. The systematic literature review  "Systematic Mapping of the Literature on Secure Software Development" \cite{9363884} analyzes various detection methods discussed in the literature. The most discussed methods are as followed:
\begin{itemize}
	\item \textbf{Vulnerability Scanning}: Vulnerability scanners are usually automated tools that check software for known security vulnerabilities. These vulnerabilities are cataloged in databases like the National Vulnerability Database (NVD) \cite{nvd}, Common Vulnerabilities and Exposures (CVE) \cite{cve}, and Common Weakness Enumeration (CWE) \cite{cwe}. These databases provide an assessment of each vulnerability. When a vulnerability is detected, the scanner generates a report that includes details and assessments from the relevant database \cite{9363884} \cite{8688018}.
	\item \textbf{Penetration Tests}: Penetration tests simulate cyber-attacks on computer systems, networks, or web applications to uncover vulnerabilities. They enable developers and security teams to identify and address vulnerabilities before they can be exploited by attackers. Automated tools can cover a wide range of attack scenarios to increase the effectiveness of these tests \cite{1392709} \cite{4402456}.
	\item \textbf{Fuzz Testing}: This method involves feeding randomly generated data to software applications to observe their response. The goal is to provoke crashes or unexpected behaviors to uncover vulnerabilities such as memory leaks, formatting errors, or buffer overflows \cite{Klees.2018}.
	\item \textbf{Static Analysis Testing}: This method is similar to vulnerability scanning but focuses on examining the code itself, without external access to the software. Static analyses are conducted while the software is not running, thus providing an effective way to identify vulnerabilities during the development phase \cite{gomes2009overview} \cite{bardas2010static}.
\end{itemize}
These methods are implemented at different points in the software development lifecycle. Static code analysis takes place during the code development phase \cite{bardas2010static}, while penetration tests are typically conducted towards the end of the development process, shortly before delivery \cite{4402456}.
\subsubsection{Response to Security Vulnerabilities} \label{theo_resp}
The second essential component in managing security vulnerabilities is their resolution, focusing on two central elements \cite{8409917}.
\begin{itemize}
	\item \textbf{Incident Response Process}:
	This process requires that a company has a well-defined plan for responding to security incidents. The process includes the following key steps \cite{torres2014incident} \cite{AD1180041}:
	\begin{itemize}
	\item Identification: Rapid detection of a security incident.
	\item Containment: Isolation of the affected system or network area to prevent further spread of damage.
	\item Investigation: Determining the causes and extent of the damage.
	\item Eradication: Repairing the damage and closing the security gap.
	\item Post-Incident Review: Assessing the measures taken to continuously improve the process and handle future incidents more effectively.
	\end{itemize}
	\item \textbf{Emergency Code Response}:
	This aspect refers to the necessity for companies to have an efficient plan in place to respond quickly in the event of a security incident. This includes the swift provision of updates or patches to fix security vulnerabilities. The plan should contain clear guidelines and procedures to enable the development, testing, and implementation of code to address security issues in a time-critical situation. The goal is to minimize the time to provide a solution for the security vulnerability while ensuring the quality and reliability of the provided fix \cite{DISSANAYAKE2022106771}.
\end{itemize}
\subsection{Security Standards} \label{theo_stand}
From the studies "Survey and analysis on Security Requirements Engineering" \cite{SALINI20121785}, "A common criteria based security requirements engineering process for the development of secure information systems" \cite{MELLADO2007244}, and "A Survey and Comparison of Secure Software Development Standards" \cite{9322704}, three standards or guidelines are particularly prominent, which are widely used in the industry.
\begin{itemize}
	\item \textbf{ISO 27001}: This standard is about the Information Security Management System (ISMS) of companies \cite{iso27}. It sets the requirements for establishing, implementing, maintaining, and continually improving an ISMS within an organization. ISO 27001 includes a list of security controls, known as Annex A, which serves as a guide for implementing security measures. These controls cover various areas such as security policies, physical security, HR security, communication security, access control, and cryptography.
	\item \textbf{OWASP}: The Open Web Application Security Project (OWASP) \cite{owasp} is an organization that is working to improving the security of software. It regularly releases resources to help developers identify and address security risks. These include software libraries, security testing tools, and documentation. An example is the OWASP Top 10, a list of the ten most common security risks for web applications.
	\item \textbf{Common Criteria / ISO 15408}: The Common Criteria for Information Technology Security Evaluation, also known as ISO/IEC 15408 \cite{iso15408}, provides a framework for evaluating the security and trustworthiness of IT products and systems. The Common Criteria specify how security features and properties of software should be defined, implemented, and evaluated. The focus is on providing a common set of requirements for the security levels of software. A central part of the Common Criteria are the Protection Profiles. A Protection Profile is a document that describes the security requirements and objectives for a category of products so that companies can select a sutable set of requirements \cite{MELLADO2007244}.
\end{itemize}

By examining these three aspects of the Software Development Life Cycle, a clear picture emerges of how security in software development is considered in the literature. In this process, clear processes, approaches, and conditions provided by guidelines and standards were identified. Thus, we can consider research question RQ2 \ref{rq2} as answered.

\section{Methodology} \label{chap:Methology}
\noindent 
In this chapter, we will look at how the interviews are structured, how the questionnaire is developed, and which method is used for evaluation. In doing so, we follow the standard work for Qualitative Studies `Real World Research: A Resource for Users of Social Research Methods in Applied Settings' by Robson \cite{robson2011real}.

\subsection{Interview Design} \label{sec:InterviewDesign}

To garner a foundational dataset for exploring existing approaches to secure software development in the industry, interviews with subject matter experts are conducted. The primary objective is to capture their impressions and viewpoints. 

Central to our data collection methodology is the autonomy of the expert respondents. It is crucial to allow them the freedom to articulate their perspectives without being constrained by a rigid set of questions, which may or may not be applicable to their specific organizational contexts and software development approaches. Therefore, we have opted for a semi-structured interview format. In this modality, participants are guided by key questions aimed at steering the conversation in a particular thematic direction. However, the semi-structured nature of these interviews allows for a conversational flow. This facilitates a more natural sharing of experiences, which might otherwise remain unarticulated if strictly structured questions were employed \cite{robson2011real}.

We have deliberately decided against using online surveys wherein participants complete questionnaires independently. Past evidence suggests that open-text fields in such surveys often remain inadequately filled \cite{robson2011real}. Furthermore, this approach would not provide the opportunity to ask follow-up questions to clarify or expand upon the participants' responses.

Each interview is planned to last between 30 to 40 minutes. This duration was determined after formulating the questionnaire, as detailed in the subsequent chapter \ref{sec:Questionary}, and subsequently validated through two pilot interviews with student colleagues. This length serves as a balanced compromise between obtaining detailed data and maintaining participant engagement. Longer interviews could potentially yield more comprehensive data, but they also risk diminishing participants' willingness to cooperate.

In selecting the participants, special attention has been paid to ensure a diverse representation across various industry sectors, thus aiming for a representative cross-section. For this purpose, we have leveraged personal contacts as well as partnerships with organizations affiliated with the academic department. These organizations were then approached to identify potential participants within their establishments. Consequently, the final selection of interviewees was facilitated by contact persons within these cooperating organizations.

\subsection{Questionary} \label{sec:Questionary}

As described in the section concerning Interview Design \ref{sec:InterviewDesign}, the interviews conducted are of a semi-structured nature. This necessitates that the questions be framed in an open-ended manner to facilitate and sustain a conversational flow. The objective is not to formulate questions that are overly rigid or prescriptive.

Subsequent to the literature review \ref{chap:Theoretical}, we have compiled a catalog of potential questions to be posed to the interviewees. Initially, we documented all questions that could be deemed relevant. These questions are designed to elicit specific pieces of information. For instance, one such question is, "How are security requirements captured in your organization?" Following this, the questions were categorized into appropriate thematic areas.

In the next phase, multiple direct questions from a single thematic block were amalgamated into an overarching question that encapsulates the entire block. The aim was to integrate these questions more naturally into the conversation and to encourage more open responses from the interviewees. During this process, we observed that some questions became overly complex and expected multiple pieces of information simultaneously. To mitigate this, we subdivided these thematic blocks into smaller segments, allocated the direct questions accordingly, and subsequently formulated new overarching questions for these smaller areas.

Through this iterative process, it became evident that certain questions were not amenable to being subsumed under an overarching question and would be more effectively posed individually. The blocks affected by this issue were those concerning the organizational context and the block titled "Current and Future Challenges." Consequently, we opted to pose the questions from these blocks directly.

All direct questions that fall under an overarching question are annotated beneath it. This ensures that the interviewer can verify during the course of the interview that all pertinent information relating to that block has been obtained. Should the interviewee fail to address any of these direct questions either partially or wholly while responding to the overarching question, the interviewer has the latitude to pose these as follow-up questions \cite{robson2011real}.

\subsubsection{Company Context} \label{question_cc}
\begin{mainBox}{Company Context}
	\begin{enumerate}
		\item What role do you assign yourself in your company with your main tasks?
		\item How many people work in your company?
		\item How long has the company been developing software?
		\item For which industries does your company develop software solutions?
		\item What type of software is it?
		\item How do you categorize the development methodology used in software development?
	\end{enumerate}
\end{mainBox}

The queries within this section aim to collect pertinent information about both the interviewee and the company they are affiliated with. This acquired data serves a dual purpose: it not only aids in the nuanced interpretation of the responses provided by individual participants but also facilitates the detection of emergent trends within the aggregate dataset. It is imperative to note that confidentiality is of the utmost importance; the collected data should preclude the identification of either the respondents or their respective companys.

For the purpose of characterizing the interviews, the following criteria have been identified as relevant:

\begin{enumerate}
	\item \textbf{Scope of Activity of the Interviewee}: Acquiring a nuanced understanding of the specific domains of software development in which the interviewee is engaged provides insights. This is especially valuable in discerning variations in security perspectives between roles, such as software architects compared to developers.
	
	\item \textbf{Organizational and Development Team Size}: This metric serves to assess whether the scale of a company influences its procedures and protocols related to software security.
	
	\item \textbf{Age of the Software Division}: This variable is gathered with a similar rationale to that of organizational size. A firm with an extensive history in software development may adopt a fundamentally different approach to issues of security.
	
	\item \textbf{Target Market of the Organization}: Sectors within the industry possess distinct security prerequisites for their software solutions. Identifying these sectors allows for an examination aimed at detecting discernible trends in the data.
	
	\item \textbf{Type of Software Developed}: Software solutions can vary significantly in their form and function, from control systems to web-based applications. Each category presents unique security challenges, and this variable aids in identifying any overarching trends.
	
	\item \textbf{Methodological Approach to Software Development}: The goal here is to distinguish between traditional models like Waterfall and agile methodologies such as Scrum. This criterion is instrumental in subsequent evaluations that seek to understand the project's development workflow in-depth.
\end{enumerate}
Moreover, the comprehensiveness of this data collection strategy is essential for ensuring that the study’s participant pool is sufficiently diverse, thereby capturing a broad spectrum of perspectives within the field.

\subsubsection{Security as Software Quality Aspect} \label{question_sqa}
\begin{mainBox}{Importance of Security}
	How would you describe the importance of security in comparison to other aspects such as functionality or user-friendliness, and how has this evolved over the years?
	\begin{enumerate}
		\item What is the importance of security compared to other qualities like functionality or user-friendliness?
		\item How has the importance of security changed over the years?
	\end{enumerate}
\end{mainBox}

This thematic block aims to investigate foundational attitudes toward security within the realm of software development. The objective is to discern the degree of importance accorded to security measures. Such information serves to provide a nuanced interpretation of decision-making processes and operational methodologies within software development practices.

\textbf{Importance and Contextual Relevance}

When combined with background information on the company, these insights may help identify potential trends. For example, the question can examine whether a company's size impacts how different aspects of software are prioritized, especially when limited developer resources are available. Similarly, questions may focus on the industry of the company’s clientele, as some sectors may place a greater emphasis on data security.

\textbf{Evolution of Security Awareness}

The second part of the question aims to explore whether there has been a change in security awareness over time. With digitalization becoming increasingly prevalent, security issues are gaining more attention in everyday life and in public consciousness. The question aims to ascertain whether this heightened awareness has correspondingly affected the importance given to security within software development.

\textbf{Customer Influence on Security Prioritization} 

An additional area of interest is whether changing customer demands have played a role in this potential shift in security prioritization. This angle is based on the idea that customer requirements might be a significant factor in shaping how much emphasis is placed on security within an organization's software development processes.

This thematic section serves to generate a comprehensive understanding of attitudes toward security by considering both organizational context and the evolving public and industry focus on security measures.

\begin{mainBox}{Security Requirements} \label{question_sreq}
	How does your company handle the topic of security requirements, from identification to implementation in software development, and how does this affect the overall development process?
	\begin{enumerate}
		\item How are security requirements captured in your company?
		\item How are they identified?
		\item When are they gathered?
		\item How do you handle it when requirements change due to new risks?
		\item How do you ensure that the security requirements are considered and implemented during development?
		\item What impact do security requirements have on software development as a whole?
	\end{enumerate}
\end{mainBox}

	The objective of this section is to meticulously examine the software development process through the lens of security, with a particular focus on security requirements.
	
	\textbf{Capturing Security Requirements: Method and Timing}
 
	The first trio of questions is designed to ascertain how these security requirements are captured. Clearly, the methodology employed for capturing these requirements is a pivotal factor in shaping the security posture of the resulting software. Additionally, it is essential to understand the temporal context within which these requirements are integrated into the development process. To offer a nuanced understanding of this, we refer back to the information on the development methodologies within the enterprise context.
	
	\textbf{Identification Strategies: Diversity of Methods}
 
	Another crucial aspect in describing security requirements is the means by which they are identified. This could entail a variety of methods, each with its own implications for the robustness of the security architecture. Hence, these questions aim to capture the spectrum of identification strategies utilized, whether it be through threat modeling, risk assessment, or other industry-specific practices.
	
	\textbf{Adaptability: Responding to Changing Requirements}
 
	The subsequent set of questions delves into the evolution of the development process. A common occurrence in software development is the changing nature of requirements as a project progresses, whether due to altered objectives or external influences like regulatory changes. To comprehend the development workflow within the company, it is vital to investigate how such changes are incorporated.
	
	\textbf{Enforcement and Verification: Mechanisms of Accountability}
 
	Following this, the survey explores how the identified security requirements are enforced and verified throughout the development process. This not only includes implementation but also involves monitoring mechanisms and checks to ensure ongoing compliance with the established requirements.
	
	\textbf{Overall Assessment: The Cumulative Impact}
 
	Finally, an evaluative judgment of the entire process is elicited. The participants are prompted to provide their impression regarding the comprehensive influence of security requirements on the development process.

\begin{mainBox}{Security Assessment and Modeling} \label{question_req_method}
	How is the quality of security for your software solutions assessed and documented?\\ \\
	Do you use specific modeling techniques for assessing and evaluating security, such as Threat Modeling?
\end{mainBox}

This section builds upon the previous segment, specifically focusing on the methods used for ensuring that security requirements are met during the software development lifecycle. It is reasonable to assume that the processes for guaranteeing compliance with security requirements would be closely related, if not identical, to the procedures used for evaluating the overall security of the software. The objective of this section is to capture these methods in detail.

Additionally, this part aims to discover whether supplementary code reviews occur through other means, such as an internal security department or external audits. Understanding the documentation process for the results of these evaluations is vital, as it could significantly influence the future direction and iterative cycles of the project.

\subsubsection{Security Guidelines and Standards} \label{question_stand}
\begin{mainBox}{Security Guidelines and Standards}
	How are security guidelines and standards for software development anchored in your company, and how do you ensure that these are considered and complied with in development projects?
	\begin{enumerate}
		\item Are there security guidelines and/or standards for software development in your company?
		\item How do security guidelines and/or standards flow into software development projects?
		\item How is compliance with security guidelines and/or standards checked?
	\end{enumerate}
\end{mainBox}

This segment seeks to explore how security guidelines and standards are embedded within the organizational structure and software development processes. Guidelines and standards can serve as the backbone of secure development, offering a framework that helps to mitigate risks and ensure compliance with legal and ethical obligations. Therefore, understanding the nature and extent of these guidelines is crucial for assessing the company's security posture.

It is also crucial to ascertain whether these guidelines are company-specific rules. To gain a deeper understanding of the process, it is essential to know how these guidelines and standards are formulated or chosen. Factors influencing this choice may include specific certifications that are granted upon compliance with these standards. It is conceivable that these certifications could be mandated by legislative bodies or required by customers.

In the case of company-specific guidelines, the intent behind their creation is also of significant importance. These guidelines may either be concrete coding practices or pertain to general approaches for requirement elicitation. The underlying intent inevitably shapes the entire software development process.

\subsubsection{Management of Security Issues} \label{question_management}
\begin{mainBox}{Internal Security Issues}
	How does your company handle security issues after the release of software, from review to deployment of a solution?
	\begin{enumerate}
		\item Is there a process for reviewing and fixing security issues after the release of the software?
		\item How are the experiences and feedback from this process used to improve security in software solutions?
	\end{enumerate}
\end{mainBox}

This section aims to explore how the company addresses security vulnerabilities that are identified or emerge after the software has been released. The focus is on examining the complete process, encompassing the aspects of vulnerability detection, assessment, team communication, and remediation. With this information, a detailed insight into the company's approach can be obtained.

Additionally, it is relevant to inquire whether an evaluation of the process is conducted after a security vulnerability has been mitigated. Such an evaluation could yield insights and potential recommendations for future practices. This directly relates to the preceding section, as modifications to internal security guidelines could result from this evaluation.

\begin{mainBox}{External Security Issues}
	What approach is there in your company regarding the use of external libraries? \\ \\
	How do you deal with security issues that arise from external influences, such as security vulnerabilities in frameworks or libraries that you use?
\end{mainBox}

This section, similar to the previous one, analyzes the handling of security vulnerabilities; however, the focus here is specifically on external code in the form of libraries. The objective is to understand the company's general attitude towards the use of external code. To investigate this, it is essential to identify the types of external code that are permitted for use. For instance, the company might allow only the use of open-source code, facilitating independent security verification. Alternatively, every library might be required to undergo security review and approval by a dedicated security department before being used. These approaches significantly influence how vulnerabilities in the code can be managed.

The second part of this section explores whether the approach to remediating a security vulnerability in an external library differs from that for internal vulnerabilities. It is expected that there will be differences in the methodology, as it may not be feasible to modify the external code.

\subsubsection{Current and Future Challenges} \label{question_challenge}
\begin{mainBox}{Current and Future Challenges}
		What challenges do you see in your work with regard to security?\\ \\
		How do you assess the status of your company in relation to these challenges and is the company prepared for them?
\end{mainBox}

This section explores what participants perceive as distinctive challenges in the realm of security. The first question aims to identify these challenges, thereby uncovering potential vulnerabilities within the company's security blueprint. By conducting this inquiry, the study could reveal whether recurring issues indicate prevailing trends, signaling industry-wide problems, especially if diverse companies have resolved similar issues within their processes.

The second question investigates whether the company acknowledges these challenges and responds appropriately. The approaches employed to tackle these challenges can provide insights into the company’s overall stance on security matters. It’s intriguing to note that when a challenge is universally recognized by multiple companies but remains unresolved, it denotes a substantial issue, warranting a more detailed exploration in subsequent studies.

In this context, examining the potential disconnect between the perceived challenges and the implemented strategies is pivotal. This discrepancy may reveal either a misalignment in organizational priorities or a gap in the comprehension and handling of contemporary security demands. Such an exploration has the potential to offer a nuanced understanding of both internal and industry-wide security landscapes and could act as a catalyst for comprehensive strategic modifications in security paradigms.

\subsection{Interview Implementation} \label{sec:InterviewImplementation}
The interviews were conducted between September 1, 2023, and October 14, 2023. A total of 21 interviews with german companies were conducted, of which 20 were usable for inclusion in the study. One interview was excluded because the company did not fit the target group. This company does not develop software itself but outsources the development as a contracted service. Therefore, it was not considered relevant for this study. In some cases, interviews were conducted with multiple individuals from a single company. However, upon reviewing the data, these were treated as separate datasets because the statements of the individuals differed significantly. This variance was often due to their employment in different departments or their positions offering distinctly different perspectives on the development process. Thus, these 20 interviews represent insights from 13 different companies.\\

The interviews were divided into three phases. Initially, we introduced ourself and briefly provided an overview of the interview's purpose and procedure. This was followed by the main interview session. In conclusion, there was an opportunity for participants to ask questions about the further course of the study and to provide feedback. The average duration of the interview phase was 29.5 minutes, with the longest interview lasting 45 minutes and the shortest being 15 minutes. The majority of interviews were conducted in the form of online meetings. Additionally, two interviews were conducted in-person at on-site locations. At the request of two participants from the same company, they were interviewed together. However, their responses are also considered separately in the analysis. Prior to the interviews, participants were provided with a consent form outlining the data usage and recording procedures, to which they agreed.\\

The recordings captured only the audio portion of the interviews. QuickTime was employed for this purpose to ensure purely local recording, rather than relying on the recording services of the meeting software. For further processing, these recordings were subsequently transcribed locally using the Whisper software, with German and English language models set to medium resolution. Given that the transcription was not flawless in every instance due to the use of specialized terminology, the transcripts were manually corrected thereafter. After verifying the accuracy of the transcripts, the audio recordings were deleted.\\

During the conduct of the interviews, several points emerged that were subsequently adjusted for the following interviews. Since these were identified early on, it was possible to capture the necessary information in all cases. However, it should be noted that not all participants were able to comment on every question. This was either because the questions fell outside their area of responsibility, or alternatively, because the question was not applicable to their company. The following are all the aspects that were modified:\\

When capturing the size of the company, it was initially overlooked in the questionnaire design that some of the companies have other business areas besides software development. This means that the number of employees does not necessarily represent the size of the development department. Therefore, in the interviews, questions were also asked about the size of the development teams.\\

Additionally, it should be noted for these companies that they might have existed longer than they have been developing software. As a result, an answer to the question of how long the company has been developing software could not always be provided.\\

In the questionnaire, in the section on security requirements with question 5, in the sections on security assessment and modeling, in the security policies section with question 3, and in the security management section, questions were asked about how security is evaluated and how compliance with requirements and standards is ensured. In designing the questionnaire, these areas were considered as separate aspects. However, participants viewed them as a single process that cannot necessarily be separated. Therefore, in the interviews, a comprehensive question was asked about how security evaluation and scanning takes place.\\

In the section on external security issues, the use of external libraries and their use of open-source were discussed. It emerged that a specific investigation into possible guidelines or rules for the use of these libraries could be interesting. Therefore, subsequent questions asked whether there is a review process or approval for the use of these libraries.

\subsection{Interview Evaluation} \label{sec:InterviewEvaluation}
As outlined at the beginning \ref{sec:ResearchQuestions}, the aim of this study is to provide insights into the methods and approaches used by companies in the realm of secure software development. Following the collection of all data, the objective now is to present the core elements, and to identify trends and possible correlations. For this purpose, we have decided to divide the analysis into three sections. This approach is based on Philipp Mayring's qualitative data analysis \cite{mayring1994qualitative}, encompassing three fundamental objectives: summarizing, structuring, and explication. These goals are implemented in the study.\\

The summary aims to reduce the contents of the data to the essentials, enabling a vivid representation of the data. In our work, this involves examining certain categories. These categories each represent a specific aspect of the data. The initial determination of the categories is based on the objectives defined in the questionnaire chapter \ref{sec:Questionary}. Thus, they are closely aligned with the posed questions, yet not limited to them. As described in the chapter \ref{sec:InterviewImplementation}), some aspects emerged as redundant during the interviews, while others, not considered initially, proved relevant. The categories reflect these changes and were thus adapted from the questions. After establishing the categories, which will be explained individually in the following chapter, codes were defined for classifying the data within each category. The process to define these codes is guided by Grounded Theory, as proposed by Strauss and Glaser \cite{strauss1994grounded}. Initially, 5 out of 20 interviews were randomly selected and examined for these categories. Relevant information was noted, and sensible groupings formed. Once codes for all categories were established, the remaining interviews were analyzed and assigned corresponding codes. If relevant information was identified in a category for which no suitable code existed, a new code was added. Subsequently, all interviews were re-examined with the revised code list to determine if a different coding would be more appropriate.\\

For a vivid representation, appropriate diagrams will be created to illustrate the distribution of the codes. It should be noted that assigning multiple codes per category is both feasible and desirable \cite{mayring1994qualitative}. It must also be taken into consideration that it is possible that participants may have omitted certain points during the interviews, for instance, if these did not come to mind at that moment. For the coding process, only those codes that were explicitly mentioned were selected. \\

The second goal of the analysis is structuring. This involves elucidating relationships and recognizing trends. In this step, the focus is on what stands out from the previously captured codes. This includes investigating whether there are overarching similarities or detectable correlations between categories. It is important to note that due to the nature of qualitative research, these cannot always be quantified statistically, as the dataset is limited. The findings of this investigation will be presented and discussed in the chapter \ref{c_Dat_Struc}.\\

The final aim of the analysis is explication. According to Mayring, additional information should be used to better explain and interpret the data from the study \cite{mayring1994qualitative}. In this work, results from the literature research, as presented in the chapter \ref{chap:Theoretical}, will be utilized. This involves examining whether the results identified in the interviews can be related to the literature data, uncovering similarities and differences.

\section{Interview Analysis} \label{chap:Analysis}
\noindent This chapter will describe all the selected categories and their respective codes, followed by the three goals of the qualitative analysis \cite{mayring1994qualitative}.

\subsection{Summary} \label{cap_Data}
In this section, the first objective of data analysis according to Mayring \cite{mayring1994qualitative} is carried out by processing and presenting the data as described in Chapter \ref{sec:InterviewEvaluation}. As noted, the categories are based on the thematic blocks of the questionnaire, but also take into account the changes described in Chapter \ref{sec:InterviewImplementation} that emerged during the interviews. We will now examine the defined categories and the codes contained within them. We identified 19 categories that can be grouped by subject.

\subsubsection{Company Context} \label{code_cc}
Firstly, we consider the questions related to the company context. It makes sense to view each of these as a separate category, as they serve to contextualize the other questions.

\textbf{Company Size}

\begin{figure}[b]
\centering
\begin{minipage}{.5\textwidth}
  \centering
  \includegraphics[width=0.95\linewidth]{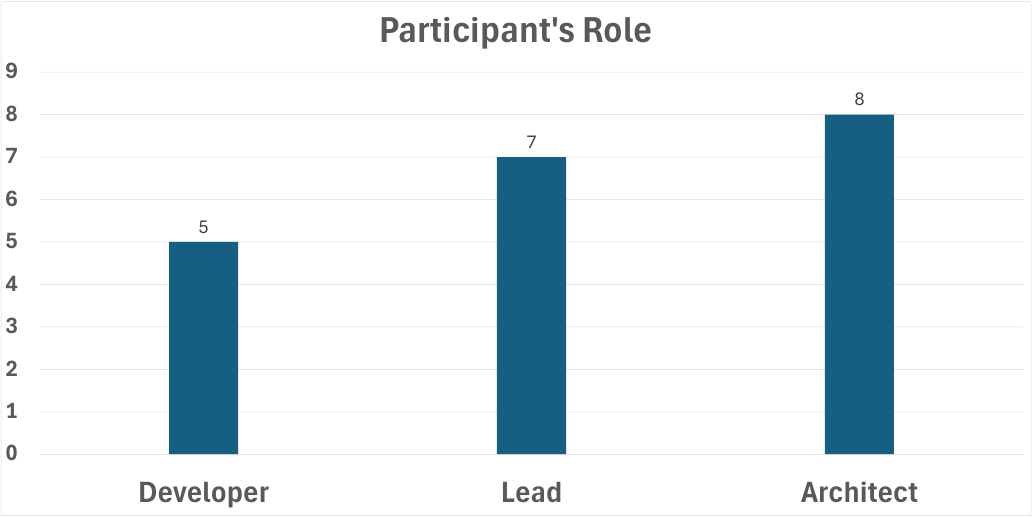}
  \captionof{figure}{Distribution for ``participant's role''}
  \label{graph_size}
\end{minipage}%
\begin{minipage}{.5\textwidth}
  \centering
  \includegraphics[width=0.95\linewidth]{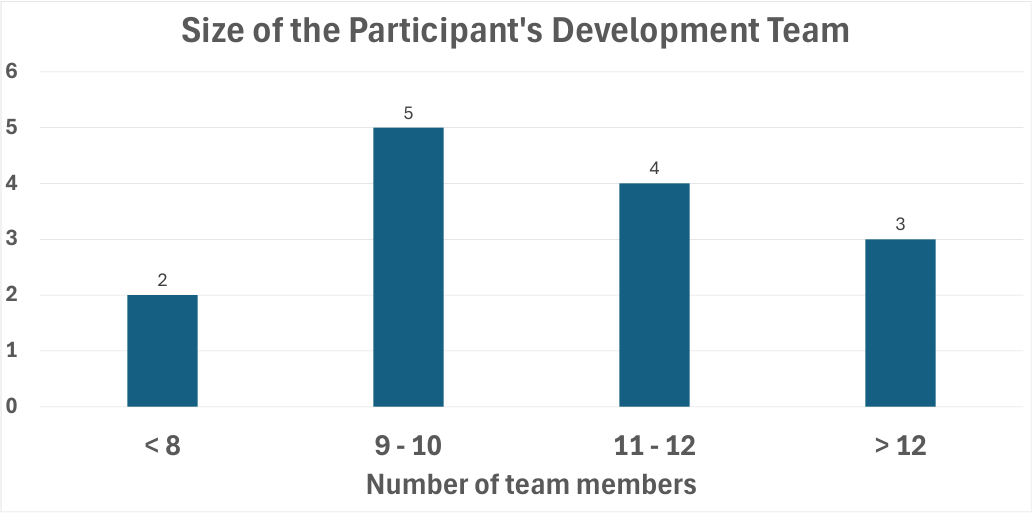}
  \captionof{figure}{Distribution for ``development team size''}
  \label{graph_team}
\end{minipage}
\end{figure}

The categorization of the company size is based on a statistical approach. To create a good distribution, the categories are oriented according to the quartiles. This results in the groupings 
\begin{itemize}
  \item \textbf{< 500}
  \item \textbf{500 - 1000}
  \item \textbf{1001 - 5000}
  \item \textbf{> 5000}
\end{itemize}
The distribution is shown in Fig. \ref{graph_size}.

\textbf{Company's Experience in Software Development}

\begin{figure}
\centering
\begin{minipage}{.5\textwidth}
  \centering
  \includegraphics[width=0.95\linewidth]{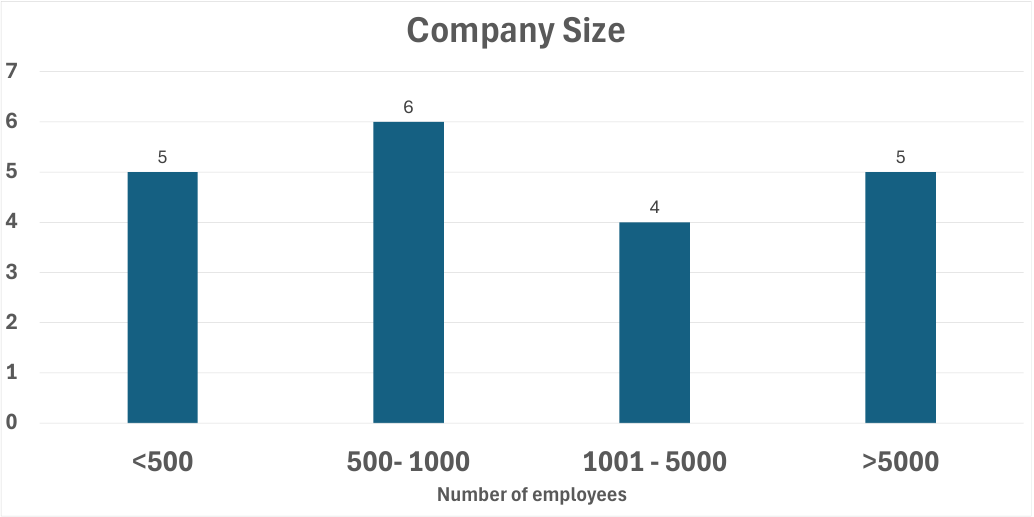}
  \captionof{figure}{Distribution for ``company size''}
  \label{graph_Kat_Role}
\end{minipage}%
\begin{minipage}{.5\textwidth}
  \centering
  \includegraphics[width=0.95\linewidth]{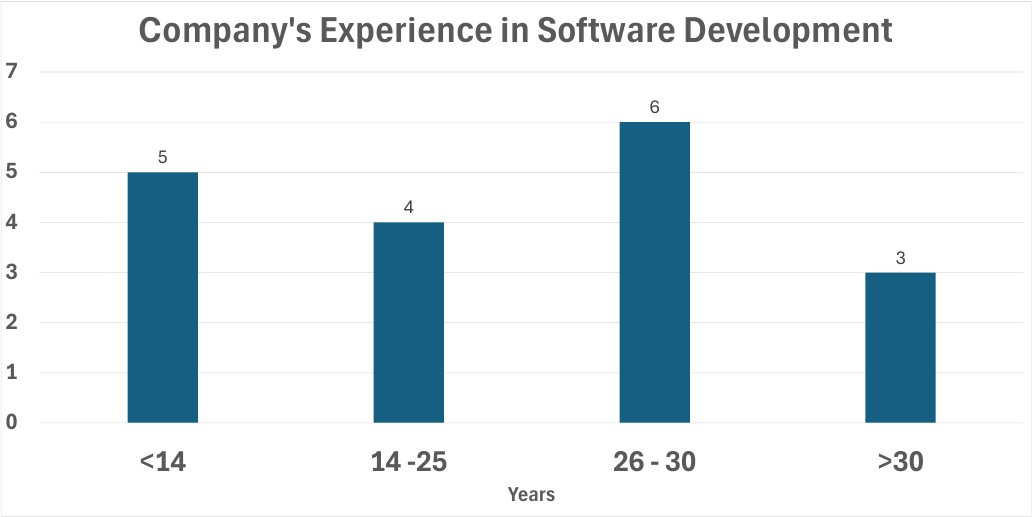}
  \captionof{figure}{Distribution for ``company's experience''}
  \label{graph_alter}
\end{minipage}
\end{figure}

This method is also applied to the question of how long the company has been developing software. Unfortunately, data is not available from all companies in this case either. The groupings are 
\begin{itemize}
    \item \textbf{< 14}
    \item \textbf{14 - 25}
    \item \textbf{26 - 30}
    \item \textbf{> 30}
\end{itemize}
The distribution is shown in Fig. \ref{graph_alter}.

\textbf{Participant's Role}

In exploring the role of individuals, our objective was to gain insight into their perspectives on software development and, consequently, on security practices. Analysis of participant responses reveals a diversity in job titles and associated responsibilities within companies. To effectively categorize these roles, three distinct codes were employed. These codes aim to delineate the degree of involvement in direct development activities.

\begin{itemize}
  \item \textbf{Development}: This code encompasses roles directly engaged in software development activities.
  \item \textbf{Architect}: This code is assigned to participants primarily involved in architectural aspects of software development, indicating a slight detachment from direct development tasks.
  \item \textbf{Lead}: Participants falling under this code predominantly assume managerial responsibilities in the development process.
\end{itemize}

The distribution of these role assignments is illustrated in Figure \ref{graph_Kat_Role}.

\textbf{Size of the Participant's Development Team }

For the size of the development teams, a similar approach is taken. Unfortunately, data is not available from all companies at this data point. This results in the groupings 
\begin{itemize}
    \item \textbf{< 8}
    \item \textbf{8 - 10}
    \item \textbf{11 - 12}
    \item \textbf{> 12}
\end{itemize} 
The distribution of these is shown in Fig. \ref{graph_team}.

\textbf{Company's Target Market}

\begin{figure}
  \centering
  \includegraphics[width=0.95\linewidth]{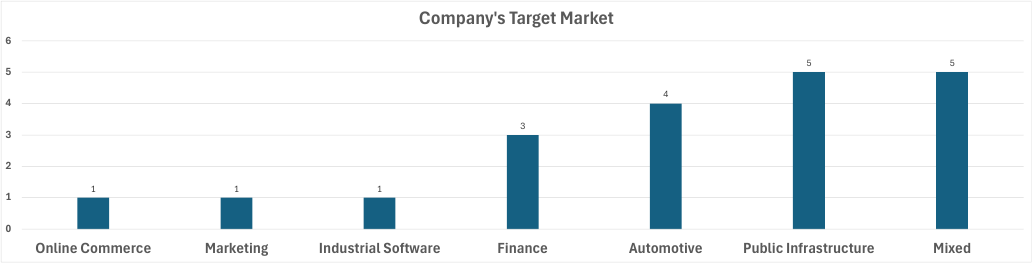}
  \captionof{figure}{Distribution for ``target market''}
  \label{graph_Kat_Market}
\end{figure}

This category reflects the response to the question regarding the industry for which the company develops software. Several sectors were identified within this category, as follows:
\begin{itemize}
	\item \textbf{Finance}: Companies in this sector develop software for the financial industry, including banks, insurance companies, and taxation systems.
	\item \textbf{Online Commerce}: In this sector, software is developed to facilitate online trading and sales. An example of such software is the development of an online shop.
	\item \textbf{Public Infrastructure}: Companies in this sector develop software for public infrastructure use. This includes the provision of public networks such as electricity, water, or gas, as well as software for public transport operations.
	\item \textbf{Marketing}: This code is assigned to companies that have described developing software for clients in the marketing sector.
	\item \textbf{Automotive}: This code denotes companies that develop software for the automotive industry, which can include both manufacturers and suppliers.
	\item \textbf{Industrial Software}: Companies marked with this code have indicated that they develop software for industrial enterprises.
	\item \textbf{Mixed}: Finally, a code was defined for companies that cannot be clearly categorized or are not specialized in one field, developing software for a broad range of companies.
\end{itemize}
The distribution of industries is shown in Fig. \ref{graph_Kat_Market}.

\textbf{Types of Software developed by the Company}

\begin{figure}
  \centering
  \includegraphics[width=0.95\linewidth]{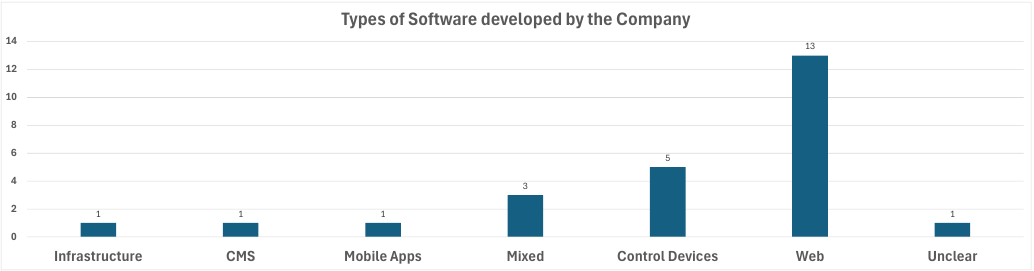}
  \captionof{figure}{Distribution for ``type of software''}
  \label{graph_Kat_Type}
\end{figure}

In this category, the type of software developed is examined, reflecting the fifth question of the company context \ref{question_cc}. A company can be assigned more than one code in this category. The following types have been identified:
\begin{itemize}
	\item \textbf{Web}: This code covers all software primarily operated through a web interface, such as a website. It is important to note that web development usually involves backend development as well.
	\item \textbf{Infrastructure}: This code represents companies that provide a platform for software operations, such as the provision and operation of Kubernetes clusters.
	\item \textbf{Control Devices}: This code refers to software used for control devices. Examples include the control of machinery in industrial plants. This does not exclude the possibility of using a web interface to access this software, which would then also be marked with the `Web' code.
	\item \textbf{CMS}: This code describes a Content Management System. No further details were provided on how this is implemented.
	\item \textbf{Mobile Apps}: This code describes the development of mobile apps, such as Android and iOS apps. Similar to `Web', a backend may be included.
	\item \textbf{Mixed}: This code is used when it is not clear what type of software is involved. Similar to the Target Market category, companies with this code are marked when they develop software for a wide range of customers and thus no general statement can be made.
\end{itemize}
The distribution of the codes is shown in Fig. \ref{graph_Kat_Type}.

\textbf{Development Process Model followed}

\begin{figure}
  \centering
  \includegraphics[width=\linewidth]{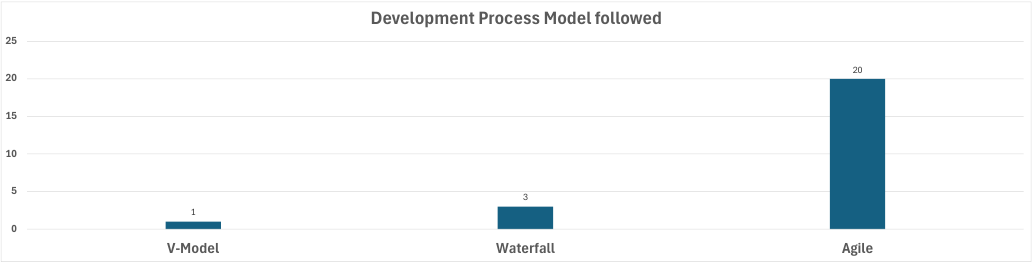}
  \captionof{figure}{Distribution for ``development process model''}
  \label{graph_Kat_Method}
\end{figure}

This category reflects question 6 of the company context \ref{question_cc}. It aims to determine the approach used in software development. Three codes were identified in this context. It is noteworthy that some companies are marked with multiple codes if various methods are employed. This is due to requirements for the development model imposed by regulations or customers for specific segments of the software. These companies noted that in the absence of specific requirements, agile methods are employed.
\begin{itemize}
	\item \textbf{Agile}: This code encompasses all agile methods. There is no explicit distinction between different methods within agile development. This is justified by the fact that almost all assignments of this code involve some form of Scrum. However, it was decided against using Scrum as a separate code because each of these companies reported using a modified form based on Scrum, rather than a pure Scrum process. SAFE and Kanban were also mentioned in individual cases.
	\item \textbf{Waterfall}: This code is used for companies where a waterfall process is in place. Similar to the agile processes, it was noted that a pure form of waterfall is not always used.
	\item\textbf{V-Model}: This code is assigned when elements of the V-Model are used in the development process. Here too, the application is not always a pure form of the model.
\end{itemize}
The distribution is as indicated in Fig. \ref{graph_Kat_Method}.

\subsubsection{Attitude towards Security}\label{code_attit}
This section examines the general attitude towards security and the shift in prioritization.

\textbf{Importance of Security}

\begin{figure}
\centering
\begin{minipage}{.5\textwidth}
  \centering
  \includegraphics[width=0.95\linewidth]{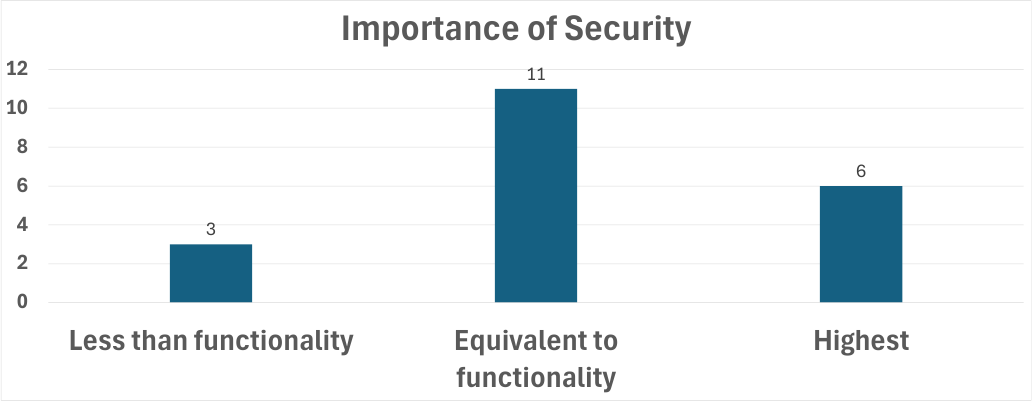}
  \captionof{figure}{Distribution for ``importance of security''}
  \label{graph_Kat_Bedeutung}
\end{minipage}%
\begin{minipage}{.5\textwidth}
  \centering
  \includegraphics[width=0.95\linewidth]{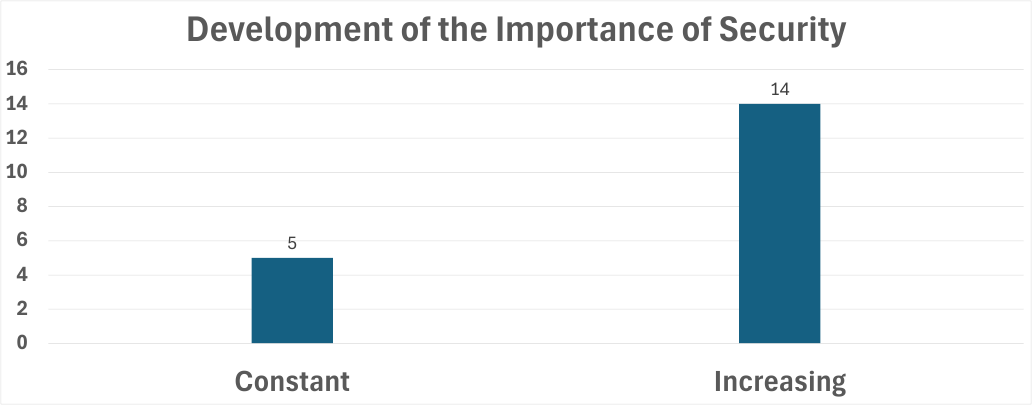}
  \captionof{figure}{Distribution for ``development of importance''}
  \label{graph_Kat_Entwicklung}
\end{minipage}
\end{figure}

In this category, the general attitude towards security within the company is reflected. This category relates to question 1 from the `Importance of Security' block of the questionnaire \ref{question_sqa}. Three different responses were identified.
\begin{itemize}
	\item \textbf{Less than functionality}: With this code, companies are coded that have stated that in case of doubt, security can be neglected in favor of feature development. An example is in interview 11, `Then it was primarily about functionality and performance, as they were not responsible for the final deployment […]. And at that time, security aspects, as we know them today, were less relevant.'
	\item \textbf{Equivalent to functionality}: This code encompasses companies that state security is highly important, but without functionality, security has no use. This means that it is considered in all developments, but possibly after functionality. Examples include interview 2, `It's always a trade-off. So I wouldn't say security is number one, but it's a close number two.' and interview 14, `So security is of course hugely important. […] Of course, you always have to weigh it against functionality and other quality attributes. Functionality plays a central role, because that is ultimately what it's about [when solving the problem].'
 	\item \textbf{Highest}: These companies view security as the top priority in software development. This means that all changes being implemented must always be considered from a security perspective. Examples include in interview 1, `And security is, I believe, the most important aspect for us. In every discussion, from conception, planning to development, testing, our risk control, that is, IT security, is always included.' and in interview 13, `So, security plays a big role for us due to the regulated environment […]. There are also, well, I would say processes or control points in our process. For example, no software that brings a significant change, or no interaction that brings a significant change, can go live without a penetration test.'

\end{itemize}
The distribution of the codes is illustrated in Fig. \ref{graph_Kat_Bedeutung}.

\textbf{Development of the Importance of Security}

This category directly follows the previous one. It reflects how the significance of security has changed over the years, thus representing question 2 from the `Importance of Security' block \ref{question_sqa}. Only two codes were identified in this context.
\begin{itemize}
	\item \textbf{Constant}: This code marks companies that have indicated the significance of security has hardly changed. It is important to note that there are two different backgrounds to consider depending on the coding of the previous category. This will be further discussed in Chapter \ref{c_Dat_Struc}. For example, in interview 13, `No, due to the regulations, the importance has always been high.' and in interview 2, `I wouldn't say that it has grown extremely. I believe it has grown tremendously in the industry as a whole. For us, it has always been of relatively high importance.'
	\item \textbf{Increasing}: The remaining companies state that security is gaining more relevance and thus also importance in development. An example is in interview 15, `Its importance has grown, especially now as regulations exert high pressure. […] In recent years, particularly in the IoT environment, we've noticed that security incidents have led to the discontinuation of entire products. So, it's definitely a topic that concerns our customers.'
\end{itemize}
The distribution is illustrated in Fig. \ref{graph_Kat_Entwicklung}.

\subsubsection{Security Requirements} \label{code_sreq}
In this section, the categories dealing with the collection of security requirements are examined. This reflects the section \ref{theo_sr}.

\textbf{Means of Security Requirements Elicitation}

\begin{figure}
  \centering
  \includegraphics[width=\linewidth]{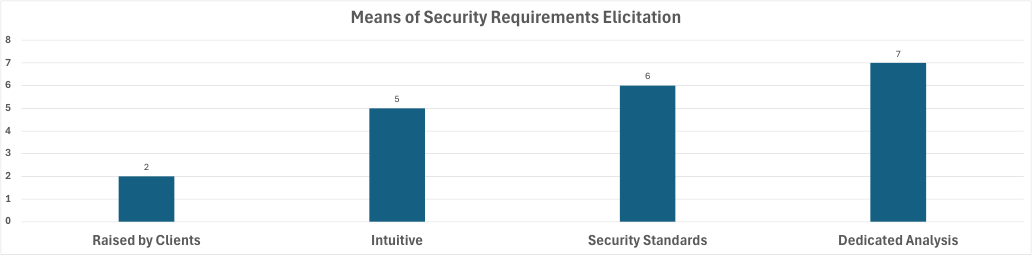}
  \captionof{figure}{Distribution for ``means of security requirements elicitation''}
  \label{graph_Kat_Analyse}
\end{figure}

In this category, it is intended to capture how companies proceed when creating security requirements for software development. This category thus represents questions 1 and 2 of the Security Requirements block \ref{question_sreq}. In this category, four different approaches were identified. Almost all companies mentioned only one method, but one noted that this can vary depending on the project.
\begin{itemize}
	\item \textbf{Dedicated Analysis}: These companies stated that they raise security requirements explicitly, e.g. through special security audits. This means that it is clearly determined what is needed and what must be implemented. For example, threat modeling is used in this process. Examples of this are in the interview 4 `If you now have a project or a new development on the project, where, for example, it is about installing a user login, then a much higher value is placed on security and on the requirements and then this answer is also more strongly documented, than if now in the further development towards the button should now be changed. So, in that sense, how we proceed is, I think, essentially that we go through different scenarios, attack scenarios, such as in authentication via OAuth, how can the token, which contains sensitive data so that you can authenticate yourself, how is it protected or how can it also be intercepted. So, we basically go there and look at what scenarios are possible, what alternatives there are to it, and what are the previous disadvantages of it.' and in interview 15 `Usually, as a starting point, you do a Threat and Risk Analysis and Risk Assessment. TARA is basically comparable to a Hazard and Risk Analysis, as done for Functional Safety topics. That is, there, risks are systematically examined, and the impact is basically estimated. So that you have actually already very early then identified what can go wrong and from this security requirements are then actually derived. And they are then basically tracked over the various phases of development.'
	
	\item \textbf{Via Security Standards}: In this approach, existing security standards such as ISO 27001 or OWASP are used as the basis for the requirements. These may be adapted to the specific software as needed. Examples of this approach are in interview 3 `Yes, so generally we decided a few years ago to have our entire development process certified according to 27001. And there are accordingly already requirements that we have to meet. And from this, essentially the entire structure of an information security management system emerges. And internally we have defined several guidelines, set our benchmark, which we must consider in our software development.' and in interview 13 `So there are also catalogs, but they are very general. There is usually not much to be gained from them, but you have to actually look closely at what I actually have to do. […] How they are identified is then simply afterwards through, in the analysis and conception phase, exactly, so a security analysis we do not make in that sense.'
	
	\item \textbf{Intuitive}: This approach describes that there is no direct process, but rather the security requirements are met by expecting active consideration from the developers. This is described in interview 18 `I believe, actually something like a security analysis, such an official process that is tied to security for the software, I don't think we have. […] I believe, at least I have not been in such a process, where people have talked, hey, listen, now let's focus on security. It is more so that we place a lot of value generally on best practices. And once a best practice is considered or once a tech stack is considered, where we simply use certain elements, libraries, at least the major ones, that we then increasingly rely on those and do not without necessity suddenly bring in other dependencies.'
	
	\item \textbf{Raised by Clients}: In this approach, the security requirements are not created by the company itself, but by the client for whom the software is intended. This is described in interview 14 `We are currently in the process of documenting central requirements, documenting them, which apply to all projects and must be adhered to. And these then serve as a basis for all projects. If there are further requirements from customers, they must then be complied with depending on the project, supplemented. Possibly they are even so good that we say, we adopt them into our central list.'
\end{itemize}
The distribution is illustrated in Fig. \ref{graph_Kat_Analyse}.

\textbf{SDLC phase of Security Requirements Elicitation}

\begin{figure}
\centering
\begin{minipage}{.5\textwidth}
  \centering
  \includegraphics[width=0.95\linewidth]{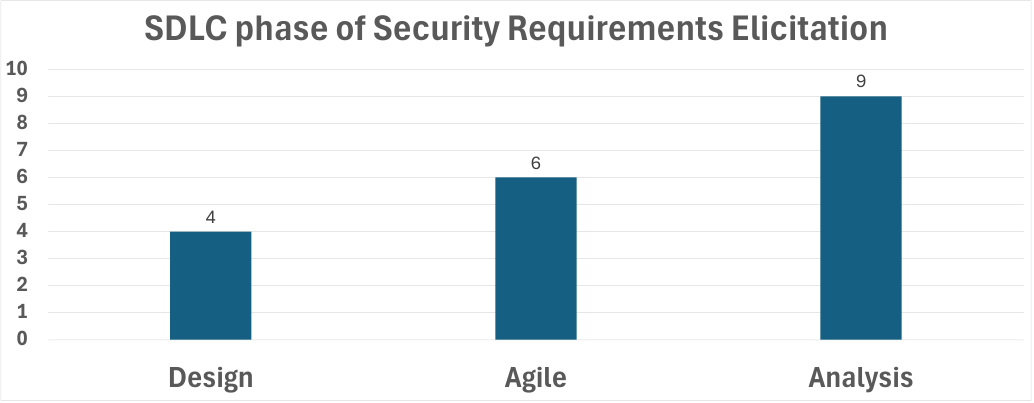}
  \captionof{figure}{Distribution for ``SDLC phase of elicitation''}
  \label{graph_Kat_Zeitpunkt}
\end{minipage}%
\begin{minipage}{.5\textwidth}
  \centering
  \includegraphics[width=0.95\linewidth]{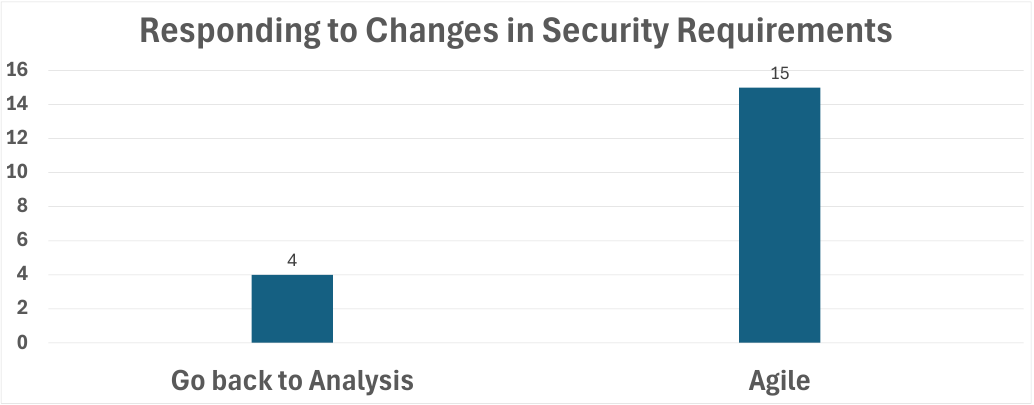}
  \captionof{figure}{Distribution for ``response to changes''}
  \label{graph_Kat_Anpassung}
\end{minipage}
\end{figure}

This category builds on the previous one. Here, it is determined at what point in the development process the security requirements are captured. This refers to question 3 of the Security Requirements block \ref{question_sreq}. Three points in time were identified:
\begin{itemize}
	\item \textbf{Analysis}: Under this coding, the collection of security requirements takes place right at the beginning of development. This means that security is already examined and the security requirements defined during the initial conceptualization of the software. An example of this is found in interview 15 `Usually, as a starting point, you do a Threat and Risk Analysis and Risk Assessment.'
	\item \textbf{Design}: In this approach, security requirements are collected once a finished concept of the software exists but before it goes into implementation. This is described, for example, in interview 4 `So, in principle, we have the development process already such that there is always a professional concept beforehand. In the professional concept, it is then roughly recorded how the general processes take place. So, the user presses the button, is redirected to the login, and so on. This is then designed by UI and UX conceptors. Once this is more or less approved, a technical concept takes place. And in a technical concept there, actually, I'd say, more or less thoroughly, depending on the use case, depending on the context you are in, such requirements and scenarios are then discussed and documented. So, it is already a step before the actual development.'
	\item \textbf{Agile}: This process ties in with the intuitive process of the previous category. Here, the requirements are established and defined over the course of the software's development. Hence, no a clear phase can be defined in which the security requirements are elicited. For instance, this is described in interview 17 `Normally, it is so that we, we call it a pitch, which is similar to preparing Scrum. It usually lasts between two and six weeks. And it is driven by content. And if we notice that this brings potential security problems, then we discuss them beforehand, before we implement it. Otherwise, it is purely content-driven. And security is, I'd say, a concern, similar to the case of performance, where the company expects to a certain extent that it is simply there. So we don't talk much about it, it just has to be implemented.'
\end{itemize}
The distribution is illustrated in Fig. \ref{graph_Kat_Zeitpunkt}.

\textbf{Responding to Changes in Security Requirements}

This category looks at the 4th question posed in the Security Requirements block. Two different methods could be identified.
\begin{itemize}
	\item \textbf{Go back to Analysis}: In this approach, a new analysis process is initiated to reassess the security investigation and redefine the security requirements. An example of this can be found in the interview 4 `Of course, you also look at it again during development if somehow requirements change or if during the development you realize, okay, now scenarios have emerged or have been noticed that were not clear before. Then you take a step back, revise the technical concept, and then continue with the development. So, it's a bit waterfall-like in that sense, that you can of course also go through interactions again.'
	\item \textbf{Agile}: In this approach, a change in security requirements is incorporated through the agile development process. Its design allows for changes in requirements. Therefore, these companies also reflect a security change in this way. This can be seen in the interview 6 `And of course, in development, you always notice something, but that's what an agile process is for.' and in interview 18 `Yes, so a normal process is initiated and prioritized. So, it's just, it's simply handled on a level with bugs and improvements, new feature developments.'
\end{itemize}

\subsubsection{Usage of Standards} \label{code_std}
In this section, the use of both internal and external guidelines and standards is examined. This reflects the aspect of \ref{theo_stand}.

\textbf{Use of  Security Standards}

\begin{figure}
  \centering
  \includegraphics[width=\linewidth]{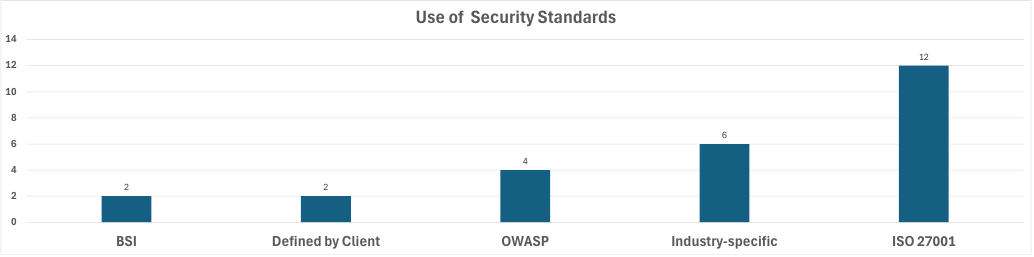}
  \captionof{figure}{Distribution for ``security standards''}
  \label{graph_Kat_Extern}
\end{figure}

In this category, it is represented which standards and regulations are applied by the companies, and in some cases must also be applied. This relates to the Security Standards and Guidelines block \ref{question_stand}. External and internal guidelines are considered separately. In this first category, the external standards are represented. Five codes were identified.
\begin{itemize}
	\item \textbf{ISO 27001}: The majority of the companies stated that they follow the ISO standard at least partially, with particular reference to standards 27001 and 9001. These provide, as described in chapter \ref{theo_stand}, a basis for information security in software development. This code includes companies that have stated that they determine their requirements based on the ISO standard, companies that align their processes with it, and companies that are explicitly ISO certified. It should be noted, that a code was only assigned with explicit mention of the standard. However, no distinction was made whether ISO, ISO 27001, or ISO 9001 was mentioned. Examples of this are in the interview 14 `So we are currently in the process of getting ISO 27001 certified. We don't have it yet. It's ongoing right now.' and in interview 19 `We are ISO 27001 certified, I believe.'
	\item \textbf{OWASP}: Another standard considered in chapter \ref{theo_stand} is OWASP. This was also mentioned by some companies. For example, this is mentioned in interview 11 `It's mainly based on OWASP. To be honest, that's enough for us for now.'
	\item \textbf{BSI}: The security guidelines of the Federal Office for Information Security were also mentioned by some companies. An example of this can be found in interview 20 `They are actually mainly derived from the requirements of the BSI.'
	\item \textbf{Industry-specific}: In addition to general standards like ISO or OWASP, some companies mentioned that they must adhere to industry-specific standards and guidelines. These include, for example, the guidelines mentioned in interview 1 `We are subject to regulatory frameworks like BAIT and MaRisk.'
	\item \textbf{Defined by Client}: Besides official standards, some companies said that they must also adhere to standards and guidelines specified by the client, depending on the project. This is seen in interview 8 `I think there are some parts of [company] that are ISO certified, which we have nothing to do with. I don't know exactly. There are also [company] guidelines, but we are not bound by them. Because we mainly do customer projects and not so much [company] internal stuff. So actually, it really depends on the client.'
\end{itemize}
The distribution is illustrated in Fig. \ref{graph_Kat_Extern}.

\textbf{Use of Internal Security Policies}

As the second aspect of the Security Standards and Guidelines block, this category considers internal guidelines. Two approaches can be found here.
\begin{itemize}
	\item \textbf{With Coding Guidelines}: This approach describes that guidelines exist within the company. These guidelines contain specific code elements that are to be applied to frequently occurring problems. Examples of these guidelines are in interview 11 `Okay, here new Java versions, this and that trick, then it obviously makes sense to supplement the guideline and add one or the other aspect and point out special situations and how they should be handled, where it's not obvious how to proceed.' and in interview 14 `That's what I meant with the centrally documented security requirements. They are to be understood as best practices, but they are also implementation rules. So it also states what you may and may not do, such as, or what is important to consider for example in identification. What also needs to be checked in JSON Web Tokens, for example, to ensure that they are valid. Such things, it's a bit of a mix of rules and best practices at that point.'
	\item \textbf{Without Coding Guidelines}: In this code, company guidelines are summarized that are formalized but do not contain code elements as examples for certain segments. Examples of this are in interview 4 `So there are, for example, requirements from the company side to only use certain tools that have been certified by our Corporate IT. Or that we rely on a particular development platform, like GitHub or something. And use the integrated tools there that are also supported by the company. On the other hand, we also have internal guidelines that we as developers have imposed on ourselves, so to speak, as best practice, which somehow meets our own quality standards.' and interview 17 `Yes, as we are a manufacturer of a wiki, we also document things accordingly. However, not always everything in the wiki, for example. So part of it is documented in the code, similar things. There we can definitely become better, no question. Otherwise, it is predominantly intrinsic motivation. So yes, we have guidelines, clearly. Some of them are also informal.'
\end{itemize}

\begin{figure}
\centering
\begin{minipage}{.5\textwidth}
  \centering
  \includegraphics[width=0.95\linewidth]{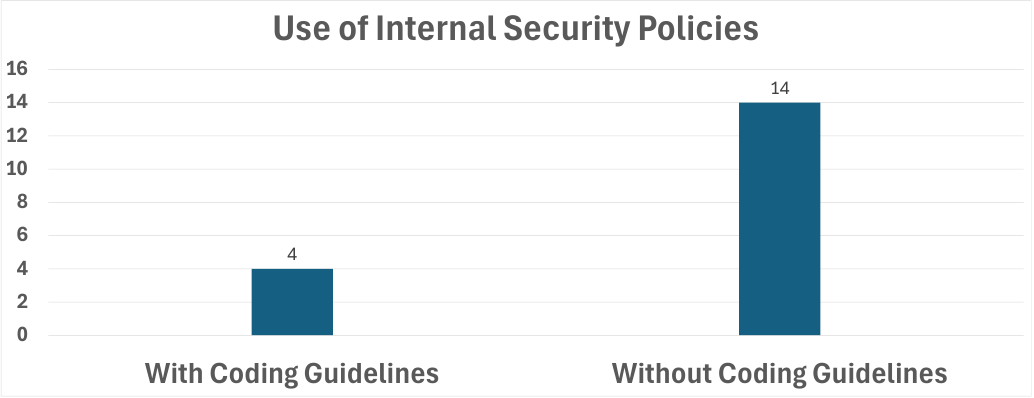}
  \captionof{figure}{Distribution for ``internal security policies''}
  \label{graph_Kat_Intern}
\end{minipage}%
\begin{minipage}{.5\textwidth}
  \centering
  \includegraphics[width=0.95\linewidth]{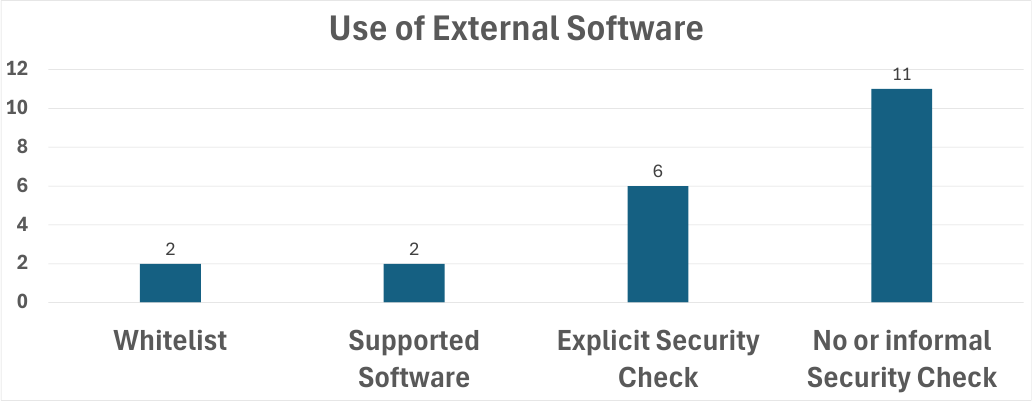}
  \captionof{figure}{codes for ``external software''}
  \label{graph_Kat_OpenSource}
\end{minipage}
\end{figure}

The distribution is illustrated in Fig. \ref{graph_Kat_Intern}.

\textbf{Use of External Software }

This category addresses how external libraries or code elements are handled in development. It should be noted that the general use of OpenSource is not marked, as this is given across all companies. The category aims to understand how companies verify the security of the libraries and thus determine which of these libraries may be used. Four relevant aspects were coded:
\begin{itemize}
	\item \textbf{Whitelist}: Companies have a defined list of libraries that can be used without further ado. This can also be realized through an existing tech stack. For such an approach, interview 9 can be considered: `First of all, internally, exactly, there is a whitelist of software or external libraries that you are definitely allowed to use, which have already gone through this certification or approval process.'
	\item \textbf{Supported Software}: In this criterion, the company requires that an external library is supported by a team of developers for the users of the library before it can be used. A third-party provider can also be set to take over this support service. An example of this is in interview 1 `And then possibly you also look for support. Most open source things are not supported. If I have a question or get a security vulnerability, it needs to be fixed quickly. So normally, any third library that we connect should be supported. […] But we definitely look for a software house or consulting firm that offers this support service. If we have questions or there, like with Log4j, a huge security vulnerability suddenly appears, then we also have a direct contact person.'
	\item \textbf{Explicit Security Check}: This approach describes that an external library must go through a verification process by the developers or a security department before it can be used in development. This approach is depicted in interview 9: `If it's now a new service that we want to connect to or a library or whatever, then it first has to go through this approval process internally from the [...] Corporate IT [to be checked]. And exactly, you submit the request and then depending on what it is, whether it's a cloud solution or, I don't know, some kind of framework or whatever, they then send you corresponding questionnaires that you have to fill out. And exactly, once you have provided all the data there, then it's also checked again, also the topic of GDPR is always an issue, especially with cloud services, if the servers are located abroad, what is then checked and then there is accordingly then a release from the central IT, whether you can use the whole thing or not.'
	\item \textbf{No or informal Security Check}: Under this code, it is understood that there is no explicit check of libraries before they can be used in development. This, of course, does not exclude that developers take a closer look at the libraries first. The security check then takes place by regularly scanning the software for known vulnerabilities. If a vulnerability is then contained in one of the libraries, it will be detected and the library can be exchanged or deactivated. An example of this approach is in interview 4 `So we mostly make sure that we use established libraries, so normally we are for example in the Springboot environment and therefore you already have quite a lot, I say, they are Open Source, but there you already have some trust that they are regularly maintained, regularly patched. We now have no specific approach to perform such a screening for example of libraries. That would probably then at most be done in the context of a code review, yes.' and in interview 19 `There is also no explicit pre-check. […] What we also do is that we have all the software that we build checked against CVEs. So that's ultimately the post-check.'
\end{itemize}

The distribution is illustrated in Fig. \ref{graph_Kat_OpenSource}.

\subsubsection{Security Vulnerabilities}\label{code_sv}
In this section, the methods companies use to identify and remedy security vulnerabilities are examined, in reference to \ref{theo_sv}.

\textbf{Security Assessment Methods}

This category deals with the evaluation and monitoring of software in relation to security vulnerabilities. It takes into account the change noted in chapter \ref{sec:InterviewImplementation} and answers the questions from the blocks of Security Requirements with question 5 \ref{question_sreq}, in the section on Security Evaluation and Modeling \ref{question_req_method}, in the section on Security Guidelines with question 3 \ref{question_stand}, and in the section on Security Management \ref{question_management}. The topic of these questions is to understand how security problems are detected. For this purpose, all mentioned methods for this detection are coded in this category. Multiple codes are specifically assigned to the companies. The identified methods are as follows:
\begin{itemize}
	\item \textbf{Code Analysis}: Potential gaps in the source code are found through checking it for certain patterns. This process is also known as static code analysis. The process of code reviews is also summarized under this code. An example can be found in interview 8 `So basically we have code reviews, so pull requests. Security plays a subordinate role there. But we are now trying to establish tools that can more forcefully promote security development. So we have just looked at Sonar Cube, that you really look at doing the code analysis more intensively or just vulnerability management.' and in interview 4 `For example, we use static code analysis that also automatically checks whether we have built in a SQL injection or not.'
	\item \textbf{CVE Scanning}: Security vulnerabilities in used external libraries can be checked by reviewing and evaluating CVE reports. An example of this approach is in interview 13 `So we are also in a containerized environment, […] we have container scanning, CVE checks, so whether the libraries you use have any known security vulnerabilities or the versions are outdated.'
	\item \textbf{Penetration Testing}: The software is tested for external vulnerabilities through penetration tests. This approach is described in interview 7 `For the cloud, we have conducted a few penetration tests. What also happens is that customers conduct such PenTests with our enterprise software. Then there are some reports that we are presented with and then we work on open issues.'
	\item \textbf{Compliance}: This code means that the company conducts tests to check whether the developed software meets all set security requirements. How these tests are carried out was not further elaborated by the companies. This approach is described, for example, in interview 1 `For example, the interface must be manually monitored every two months. It varies depending on the type of risk. You define these measures. You also define a control. Then you check at predefined intervals. This is called re-certification with us, a strange word. This means that at a predefined time, one must always manually or automatically check whether the risk occurs. Whether the measures were correctly attacked and so on and so forth.'
	\item \textbf{Threat Modeling}: Through an analysis of the functionality and architecture of the software, potential vulnerabilities are identified. This approach is mostly used for an initial assessment of security to determine security requirements. An example of this is found in interview 15 `Usually, as a starting point, you do a Threat and Risk Analysis and Risk Assessment. TARA is basically comparable to a Hazard and Risk Analysis, as done for Functional Safety topics. This means that risks are systematically examined and the impact is basically estimated. So that you have actually already very early then identified what can go wrong and from this, security requirements are actually derived.'
	\item \textbf{External Audits}: The security of the software is externally reviewed by a third party, and the results are then returned to the company. Various tests can be conducted by the external examiners. This is described in interview 12 `I think that it was also, so the security testing topic, it is not done only by QA team, in-house QA team, but I know that there is a practice to involve external companies to conduct independent security audit.'
	\item \textbf{SBOM}: Software Bill of Materials refers to a list of used code, libraries, and software. Once this list is created, it can quickly and definitively determine what is contained in the product, e.g., a container image. This allows for a fast and efficient scan for known security vulnerabilities, as seen in interview 18 `With the libraries and CVE stories, it is so that we now collect Build-of-Materials. And that means if the CSO says or finds something, then he can also look up himself, search the database, which products in which versions use it and can estimate the impact a bit.'
	\item \textbf{Security Departments}: Some of the participants are not involved in the security testing in their companies and therefore could not provide detailed information on the exact process. In these cases, the review is taken over by a security department within the company. An example of this is interview 9 `Then there are quite clear role distributions with us, that this then lies more in operation and in infrastructure. Should there be, for example, also somehow an incident or a hacker attack or a DDoS attack or similar, then these colleagues are also always directly involved.'
\end{itemize}

\begin{figure}
    \centering
    \includegraphics[width=1\linewidth]{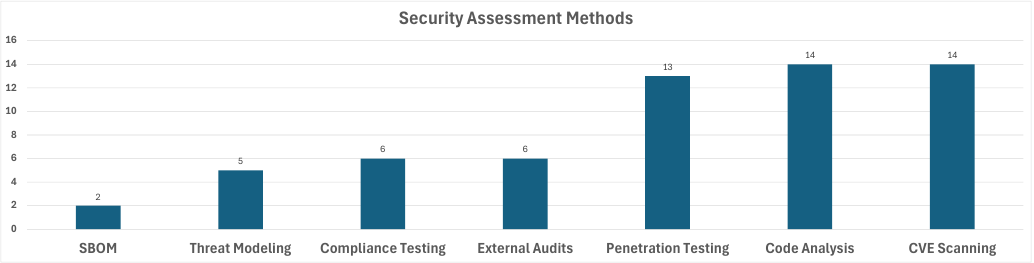}
    \caption{Distribution for ``security assessment methods''}
    \label{graph_Kat_Scanning}
\end{figure}

In addition to the aforementioned codes, ten companies explicitly mentioned that these tests are carried out automatically and eleven mentioned that tests run repeatedly. However, as the execution of the tests does not fall into this category and our data does not clearly indicate which type of tests are automated and repeated, we have not included these codes in the distribution. The distribution is illustrated in Fig. \ref{graph_Kat_Scanning}.

\textbf{Measures in Incident Handling}

\begin{figure}
    \centering
    \includegraphics[width=1\linewidth]{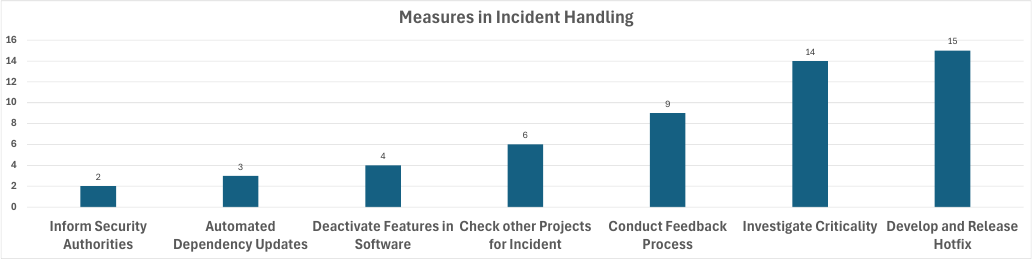}
    \caption{Distribution for ``measures in incident handling''}
    \label{graph_Kat_Update}
\end{figure}

In this category, the handling of security vulnerabilities is addressed. It covers the areas of internal and external security vulnerabilities in the Management of Security Issues block \ref{question_management}. The previous category already dealt with how these vulnerabilities are detected, so this one focuses only on the process of closing these vulnerabilities that follows. A general approach is followed, but there are differences in details. Therefore, the general procedure is first explained, and the coding only concerns the details in which the companies differ. It is also important to note that a coding was only given if it was explicitly mentioned.
The general process is a very simplified representation, where the discovered security vulnerability is fixed by adjusting internal code, updating external libraries, and possibly adapting their interfaces. Then this is played out in an update to the customers or taken over by operation in the case of Software as a Service. The differences are coded as follows.
\begin{itemize}
	\item \textbf{Investigate Criticality}: The team responsible for fixing the vulnerability first checks its impact and whether it needs to be closed immediately or can be resolved through a regular update. An example is the investigation of incoming CVE reports, as in some cases the affected functions that triggered the report are not used in the software, even if the library is used in another capacity. An example of this is in interview 3 `That means, the teams, if a vulnerability was reported, also partly automated by the tools that found it. And there the own assessment is stored. If a vulnerability, which was classified as critical by the original author, then after the own assessment is perhaps no longer critical because the functionality is not used or was intercepted by another measure or or or, then this is accordingly stored there.'
	\item \textbf{Develop and Release Hotfix}: Vulnerabilities that have been fixed can be resolved directly through a non-planned update by a hotfix, without remaining open until the next regular update. This can be particularly useful, for example, if only an update is issued every 6 months. An example of this is in interview 9 `Try to fix it as quickly as possible. And even if we only plan two or three major releases, we always have the possibility to play out hotfixes, for example for the app.'
	\item \textbf{Deactivate Features in Software}: In some cases, it may be necessary to immediately prevent a security vulnerability from being exploited before a permanent solution is available. In these cases, individual functionalities can be deactivated up to the entire software. An example of this is in interview 1 `Even we have to, in this case, if it's really a very, very big problem, then we probably have to immediately take the system off the network.'
	\item \textbf{Check other Projects for Incident}: This code denotes that the process for solving the problem involves checking other projects of the company to see if this problem also occurs there. See for this in interview 18 `And then the question is, wait a minute, was this somewhere else? [...] And is it installed anywhere else? And do we have to update it there too and send out some information that please update. With the libraries and CVE stories, it's like that we now collect Build-of-Materials. And that means, if the CSO says or finds something, then he can also look himself, search the database, which products in which versions use it and can estimate the impact a bit.'
	\item \textbf{Automated Dependency Updates}: This code describes the approach that, as part of the prevention of security problems, automatically updates dependencies in the software to the latest version. All companies that have stated this use Renovate as in interview 17 `Yes, we try to update them. We use, for example, Renovate. And what was the other thing we use? Ah, from GitHub. So we use GitHub as a hosting service. And GitHub has, I think, its own thing to open issues, to update things.'
	\item \textbf{Inform Security Authorities}: Some companies have stated that in the event of a serious security incident or attack, there may also be contact with the police and the data protection authority. This is stated in interview 16 `But then the state data protection officer must be informed. This is again a reporting obligation. Just like data protection, as soon as personal data is affected somewhere, this gap must be reported. And then it depends on how big this gap is. One can also say that the data protection might also take another look at it.'
	\item \textbf{Conduct Feedback Process}: This approach denotes that after a vulnerability is found and fixed, an analysis of the incident takes place. This can involve examining how it was handled and where the process might need to be revised. It can also involve adjusting the internal guidelines for code development to prevent similar problems in the future. This is described in interview 2 `And finally, hopefully a mitigation is performed and then retrospectively through the Post Mortem looked at what has gone wrong, but after the thing is halfway fixed, what so is patched up, then it is looked at how we can fix it long term and what all went wrong. [...] The retrospective is, we look at what happened, what was good, what was bad, and the learnings from that are then packaged into stories by the security team and then integrated into the process afterwards.'
\end{itemize}
The distribution is illustrated in Fig. \ref{graph_Kat_Update}.

\subsubsection{Challenges} \label{code_chal}
In this section, the challenges that companies perceive in relation to security and whether they feel prepared for them are examined.

\textbf{Current and Upcoming Challenges}

\begin{figure}
    \centering
    \includegraphics[width=\linewidth]{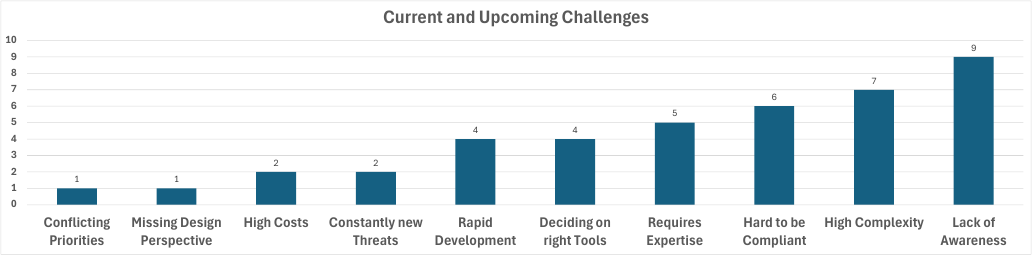}
    \caption{Distribution for ``current and upcoming challenges''}
    \label{graph_Kat_Challenges}
\end{figure}

In this category, the last block of the questionnaire Current and Future Challenges is addressed. Here, the challenges that are seen are initially coded. The following challenges have been identified:
\begin{itemize}
	\item \textbf{High Complexity}: It was frequently mentioned that security in software development is a very extensive topic with many aspects to consider. Examples of this are in the interviews interview 8 `The big challenge is that it is so complex and has so many aspects. That as a developer, you can't really handle it all on the side.' and interview 13 `It's just the rapidity and also the breadth of things that you have to deal with. And you can't really be so deeply involved in every detail that you always catch everything.'
	\item \textbf{Rapid Development}: This challenge is directly related to the previous one. It's about the field of security developing at a very high pace, requiring constant refreshing of information. An example is the already mentioned quote from interview 13 and interview 14 `I think it's a very moving target, the whole topic. So security is a complex topic and it's very dynamic. I think those are the challenges. There's a lot of movement in it.'
	\item \textbf{High Costs}: Another challenge mentioned is that comprehensive security development consumes time and thus money. An example of this is given in interview 8 `The most difficult thing I see is that if you really take it very seriously, it becomes a bit more expensive because it's sometimes quite elaborate, and finding the right mix between cost and really necessary security.'
	\item \textbf{Requires Expertise}: Building on the challenge of complexity and rapid development is the lack of appropriately skilled personnel in companies who are familiar with the subject and can conduct the necessary security assessments, as described in interview 17 `And also, especially regarding training in security concerns. So ultimately, we currently don't have a single expert in that area in our company. These are all things that we have somehow acquired on the side. And yeah, it would probably be quite good to have someone who deals with it full-time and is part of the team and ensures during development, for example, that we consider such things.'
	\item \textbf{Lack of Awareness}: Building on this, it was made clear that there is a lack of awareness about the topic among both employees and customers. This is described, for example, in interview 11 `Yes, so the biggest challenge is the awareness that needs to be created among people, and that must lead to a willingness to understand that this is a topic that everyone needs to care about. Yes, because it's not easy, people already have all sorts of things to do.'
	\item \textbf{Hard to be Compliant}: This code describes that developers often find it difficult to develop software while simultaneously meeting all security regulations and requirements, as these are sometimes set too strictly. This is shown through interview 10 `What's a challenge is complying with things that are super strict, so sometimes we can't be as agile, we can't bring things to market as quickly as other companies, because we can't afford to have a car attacked or something. Exactly, and also with us, with this product that is internal, sometimes I feel that these guidelines are actually too tight on data centers. So everything that was done on-premise, one tries to do the same in the cloud and sometimes it's super difficult, if not completely impossible to comply. Then you have a finding, but you also have no solution. You ask the architect himself, okay, but actually, we can't do this in the cloud. How would you do it? And then the answer is, well, the cloud provider should do this and that. And then you make a requirement to the cloud provider and hope that they respond to it, but sometimes they just laugh in your face and say, it's not possible.'
	\item \textbf{Deciding on right Tools}: This code means that companies find it difficult to find the right tools and aids for their processes. They justify this by saying that there are a multitude of options available, but it is not always clear what is suitable for them. An example of this is in interview 4 `So I think a key challenge is finding a suitable process. So, yes, security in development, one thing is having a good tool stack, having good tools with which one can work, especially tools that integrate well. For example, it's no use to me if I have a tool that checks my dependencies, if it then gives me an HTML report that I have to go through manually. So you also have to have a certain integration.'
	\item \textbf{Constantly new Threats}: This code summarizes that there are always new ways to attack software. This results partly from increasing complexity, as well as from the human factor. This is described in interview 14 `New attack vectors are constantly being added. Now probably AI is increasingly playing a role, making everything even more difficult. Then, of course, there's the human factor. There are many attacks that don't directly target security vulnerabilities in software but recognize the human as a weak point and then exploit the employee through phishing mails or social engineering or such variants to then insert some malicious code.'
	\item \textbf{Conflicting Priorities}: This code represents a conflict between the requirements for the functionality of the product and its security. This is described in interview 15 as follows: `And the problem is that cybersecurity requirements may be contradictory to safety requirements. For example, from a safety standpoint, I want the door to open when the vehicle is stationary so that someone can get out of a burning vehicle. From a cybersecurity perspective, that may be exactly what I don't want.'
	\item \textbf{Missing Design Perspective}: This challenge describes that currently security is not sufficiently considered in the design phase of software development. This is brought up in interview 18 `Especially in system integration and overall architecture. We have many models of the system, currently two or three, I'd say. One is a data flow level, where it states with which protocols different parts of the suite exchange information over the transport domain. Then we have one, which is simply from our point of view, what is running in the system and where is it running? Is it running as a container on Windows or Linux or something? And then we have from our point of view a network perspective, where it states, okay, we have different network partitions. […] And that's something we now consistently carry through, that we always draw or sketch something like this. And I think in addition to that, there needs to be more, there is partly security in it, but only one perspective on security, for example, the encryption of a connection. And many of these perspectives are simply missing.'
\end{itemize}
The distribution is illustrated in Fig. \ref{graph_Kat_Challenges}.

\textbf{Estimated Level of Preparation of the Team or Company to meet the Challenges}

In the final category, it is depicted how companies assess their level of preparation to meet these challenges. All participants mentioned, that there are current efforts in their teams or companies to improve their handling of security. Hence, we built codes based on their general impression on how well prepared their teams and companies currently are. 
The categorization is divided into the following codes:
\begin{itemize}
        \item \textbf{Not well prepared}: The participant does not consider his team or the company as well prepared due to manifold reasons. An example is given in interview 19: `So we are preparing for that. We are not yet well prepared.'
	\item \textbf{Somewhat prepared}: Participants' statements were assigned to this category if they indicated that although there is a certain level of preparation in the team or company as a whole, the overall impression is that there is still a lot to improve. Interview 14 gives a good impression of statements that fell in this category: ’I would say we are working on it, but of course, there are always places where you need to work on [\dots] We are currently working on the ISO 27001 certification. I think this will raise many questions for us. We will hopefully clarify many questions and thereby be even better prepared for the future. But it’s still work. Accordingly, yes, we are on it, we are working on it.’
	\item \textbf{Fairly well prepared}: We classified all statements in this category if the overall tone was better than that of the ``Somewhat prepared'' class but participants still see room for improvement. Interview 11 is a good example of statements in this class: `I think that we have taken up the matter quite well. At the same time, I also see it as an ongoing challenge. It’s not a task that you tackle once with a concerted effort and then set aside and say, okay, now it’s running, but it’s something that needs to be continuously developed further.'
	\item \textbf{Well prepared}: In this code, the participant's statements lead to the conclusion that the perceived level of preparation is overall good and it did not include a clear statement that they are still improving in this matter. Interview 1 was the only one we classified with this class: `With reservations, I think so, because we don't work alone. We work together with the provider and with our data center operator. For example, all the penetration tests or system integration tests are usually carried out by the provider. And we usually have to write the concepts on how to test the system. And yes, I would say that we are already prepared for this, but it always involves effort and costs.'
\end{itemize}
 
\begin{figure}
    \centering
    \includegraphics[width=\linewidth]{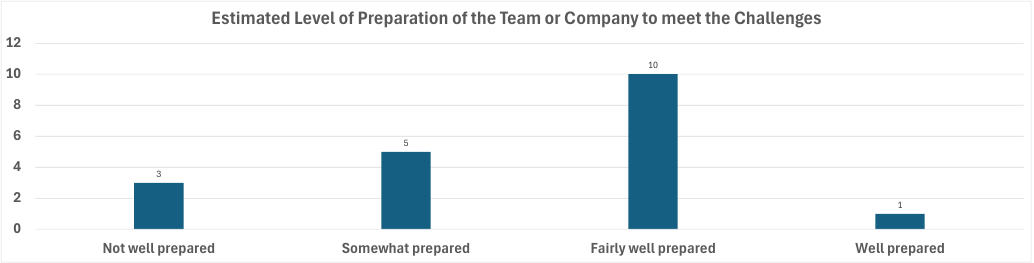}
    \caption{Distribution for ``level of preparation to meet challenges''}
    \label{graph_Kat_Ready}
\end{figure}

The distribution is illustrated in Fig. \ref{graph_Kat_Ready}.

\subsection{Structuring} \label{c_Dat_Struc}
In this section, the third research question is answered by filtering out trends and correlations from the data described in chapter \ref{cap_Data}. Due to the limited amount of data, it is not possible to find statistically verifiable correlations, but some patterns can be recognized.

In designing the questionnaire, an attempt was made to find a correlation between these characteristics and the handling of security through the corporate context. After conducting the study, we can identify the following points. Contrary to expectations, there seems to be no direct correlation between the size and age of the company and its handling of security. However, what does have an expected correlation is the industry in which the company operates. Companies whose software is used in critical areas, such as finance, infrastructure, and device control where the security of life plays a role, pay more attention to the security of their software. A quite direct indicator of this is the category `Importance of Security' \ref{code_attit}, as all companies marked there with the code `Highest Priority' operate in one of these areas. Another indicator for this assumption is that in these companies a more formalized and clearer process for the creation of secure software exists. This can be seen in the categories dealing with the collection of requirements \ref{code_sreq}, as well as the evaluation of security \ref{code_sv}.
This can also be interpreted from the development model, as more rigid methods such as the waterfall model were mentioned more frequently in these companies. However, it should be noted that the companies state that these approaches arise from regulations. Therefore, these are also an indicator, but the approach itself does not create a stricter handling of security.

A general trend that can be observed is that none of the respondents indicate that security has a low importance. While it is sometimes stated that functionality takes precedence over security in development, security is never neglected. This is also reflected in the statements on the category `Importance of Security' \ref{code_attit}. Another trend in this context is that all participants state that the importance of security has increased over time. When considering the categories `Importance of Security' and `Development of the Importance of Security' \ref{code_attit} together, it is noticeable that in almost all cases where the coding `Constant' was used for development, this is based on the fact that the companies already give the highest priority to security and therefore cannot increase it further.

When looking at the security requirements elicitation, it is noticeable that in fewer cases than expected a detailed analysis of the specific security requirements is conducted. When such an analysis is carried out, meaning in the category `Means of Security Requirements Elicitation' the code `Dedicated Analysis' was assigned, it can be seen in the category `SDLC phase of Security Requirements Elicitation' that in most cases this is done right at the beginning in the analysis phase. There is also a correlation that when security requirements are intuitively raised by developers, they are identified in the ongoing process, marked by `Intuitive' and `Agile'. The identification of security requirements by using security standards is unexpected. This is because the ISO 27001 standard, which was the most frequently mentioned security standard being used, gives very generic recommendations, as explained in chapter \ref{theo_stand}. The use of ISO 15408 would be expected, as it is specifically designed with the SREP in mind, as described in \ref{theo_srep} as well \cite{iso15408}. Additionally, many companies stated that they base their process on the ISO 27001 standard, as depicted in the category `External Standards' \ref{code_std}. From this, it can be deduced that this standard has a very high importance for development in general, as also shown in interview 16: `Then you have ISO 27001. Without it, you're out of luck, if you're not certified, you already have a problem.'

Regarding the change of requirements, it is noticeable that, as depicted in the category `Responding to Changes in Security Requirements' \ref{code_sreq}, in most cases the interactive process of agile software development is used. It is noteworthy that even companies that practice a strict and highly regulated approach to security in software development such as interview 10 benefit from this. While this company conducts a dedicated and detailed security analysis from the beginning, as can be seen from categories `Responding to Changes in Security Requirements' and `Means of Security Requirements Elicitation' \ref{code_sreq}, it responds to changes in the sprints. It has also emerged from other companies that security requirements are treated the same way as bugs or feature requirements. This suggests that security requirements should no longer be seen as a non-functional quality characteristic and requirement but, given the steadily increasing importance of security, should also be treated with the same priority as functional requirements.

Regarding the `Use of External Software' \ref{code_std}, a correlation can be seen between the importance of security \ref{code_cc} and the process of a dedicated security review of these before their use. Companies that have marked security as their highest priority often have this process implemented. However, this correlation is not a strict causality as some companies indicated the highest priority but still use no or an informal security assessment process for external software, for instance in interview 20 and vice versa companies that conduct explicit security checks but do not classify security as their highest priority as in interview 9.

Looking at the evaluation and monitoring of security \ref{code_sv}, it is immediately apparent that the most commonly used methods are code analysis, CVE scanning, and penetration tests. From the process descriptions of the interviewees, it emerges that code analysis is primarily used for self-developed code, CVE scans for external libraries, and penetration tests for access security. All companies for which the interviewee could make a statement on this topic, or where this review is not carried out by an external department, use at least one of these approaches. In general, it is noticeable in the formulations of the participants that many initially referred only to external libraries when asked about handling security vulnerabilities and less frequently, or only upon further questioning, mentioned vulnerabilities in their own code. This suggests that greater awareness can be created in this area.

Another trend in security evaluation is that all companies are trying to improve, expand, and more clearly define this process. A factor that plays a crucial role here is the automation of tests. It is a goal pursued by the participants to automate as many of the tests as possible if this has not already been done.

This aspect of automation can also be observed in handling security vulnerabilities \ref{code_sv}. Efforts are increasingly being made to automate the classification and evaluation of the resulting reports and apply them to the respective projects so that developers have a suitably individualized report available.
An unexpected observation in the remediation of security vulnerabilities is that less than half of the participants have indicated that their companies operate a feedback process, which is marked with the code `Conduct Feedback Process' in the category `Measures in Incident Handling' \ref{code_sv}. This is unexpected because it is prescribed by the ISO 27001 standard \cite{iso27}.
Furthermore, there seems to be a correlation between the coding `Deactivate Features in Software' and the priority of security. In every case where this code was assigned, it is also indicated in category `Importance of Security' \ref{code_attit} that security has the highest priority.

Regarding the challenges in software development \ref{code_chal}, it can be observed that companies see the `Lack of Awareness' and `High Complexity' as the biggest challenges. Here, the coding for `Rapid Development', and `Requires Expertise' also play a role, as they can be considered symptoms of or reasons for complexity. It is noticeable that these problems are recognized by all companies, regardless of the importance of security or the size of the company. An interesting aspect is that all companies have stated that at least they are aware of this problem and that work needs to be done to solve it.

A general observation that applies to all categories outside the corporate context is that the participants tend to view the processes and approaches of the company as more informal, the closer they are to the actual development of the software, as coded by the `Participant's Role' \ref{code_cc}.

With the help of the two chapters \ref{cap_Data} and \ref{c_Dat_Struc}, a comprehensive overview of the state of secure software development in the industry can be provided. Recurring practices have been identified, from the capture of security requirements to the remediation of emerging security vulnerabilities. As a result, research question RQ3 \ref{rq3} is answered.

To give an answer to RQ4 \ref{rq4}, our data shows a trend towards the adoption of the ``security by design'' paradigm in the industry. In category \ref{code_sreq} around half of the participants stated that security requirements are systematically raised and managed in the analysis phase of a development project as seen in Figure \ref{graph_Kat_Zeitpunkt}, hence at the beginning of the SDLC. The security requirements are furthermore raised by $65\%$ of the participants in a dedicated manner or by following and adopting certain security standards as seen in Figure \ref{graph_Kat_Analyse}. 

However, there is still much room for improvement. While the picture being drawn by how security requirements are raised and managed supports this trend, the analysis of our data regarding how security assessments are done in category \ref{code_sv} draws a slightly different picture. The only data that allows a possible interpretation of security assessments conducted early in the SDLC is the $25\%$ of participants who stated that threat modeling is applied in their projects. While the six participants who indicated that security assessments are outsourced to their security departments also allow possible security assessments to be done early on, it is unclear what kinds of security assessments the security departments actually do. However, even if we include these four responses in the proportion that do security assessments early on in the SDLC, it is still roughly $50\%$ of project teams and organizations surveyed that possibly do so. Moreover, none of the other data allow any interpretation that the system design is systematically checked for security. 

To conclude, in the early phases of the SDLC, the emphasis in the industry's pursuit of the ``security by design'' paradigm appears to be more on systematical security requirement elicitation and management rather than on systematically designing secure software systems as well as assessing the security of software systems on a design level.

\subsection{Explication} \label{c_explication}
After the research questions RQ2, RQ3, and RQ4 from Section \ref{rq2} could be answered, an answer to the RQ1 can now also be given. For this, we look at the differences and similarities between the processes and approaches of the literature and those applied in the industry in this section. For this purpose, we will discuss the three central aspects of the SDLC, which were identified in the chapter \ref{chap:Theoretical}.

\subsubsection{Collection of Security Requirements}
In examining the collection of security requirements, it becomes immediately apparent that none of the interview participants named any formally defined approaches such as SQUARE \cite{square} or SREP \cite{MELLADO2007244}. However, the activities that make up these approaches were mentioned individually. Thus, from the literature, we have identified three aspects: `Threat Modeling', `Risk Analysis', and `Requirement Review'. Threat Modeling was mentioned by five of the companies \ref{code_sreq}. The process of risk analysis in the collection of security requirements was on the other hand not explicitly mentioned by any of the participants. However, it should be noted that seven of the companies said that an analysis is conducted for the security requirements elicitation. Since no more detailed information is available on what exactly is done in this analysis, it is possible that some of the companies perform Threat Modeling and risk analysis in this step but have not mentioned it. Without further investigation in this direction, no statement can be made on this.

The `Requirement Review' can be directly identified in the statements of the participants. Thus, the approach of the implicit capture of security requirements is precisely this procedure. However, where they differ is that the Requirement Review in the literature is conducted in the context of an extended process like SREP. This also includes elements of Threat Modeling and Risk Analysis. However, from the statements of the participants, it cannot be concluded that these elements are also carried out in the implicit process. Another difference between the literature and practice is which collections of requirements are used for the Requirement Review. OWASP is mentioned in both cases, but they differ in the ISO standard used. The participants stated in the survey that the ISO standard is used as the basis for their security assessment. However, it was not clearly stated which ISO standard is meant. In the further course of the interviews, ISO 27001 is increasingly mentioned, suggesting a connection. The standard CC/ISO 15408 used by SREP \cite{MELLADO2007244} was not mentioned in any of the interviews.

\subsubsection{Handling of Security Incidents}
The approaches to the detection and remediation of security incidents also coincide in most cases between theory and practice. Thus, three of the four most commonly used methods for detecting vulnerabilities in the industry are also the most used by the interviewees. This includes `Vulnerability Scanning', `Penetration Testing', and `Static Analysis'. A clear difference, however, is Fuzz Testing, which is often considered in the literature but is not conducted by any of the interviewed companies.
In the remediation of security problems, clear parallels can also be discovered. Thus, the Incident Response Process prescribes a classification and investigation of the incident to remediate it and a final review of the approach. Both of these elements can also be found in the interviews. The remaining parts of the process were not discussed in the interviews.
Also, that a solution for the security vulnerability should be provided as quickly as possible in the form of an update is practiced by 15 of the companies.

\subsubsection{Security Standards and Guidelines}
Two of the three most common security guidelines and standards considered in the literature were also mentioned by the interviewees, these are the OWASP guidelines and ISO 27001. CC/ISO 15408, which plays a particularly important role in the literature in the collection of security requirements, was not mentioned by any of the participants.\\

The central research question RQ1 \ref{rq1} can thus be answered that the representation in the literature applies to practice, although it is significantly more formal than in practice due to the nature of scientific work.

\section{Conclusion} \label{chap:Conclusion}
\noindent This last chapter provides a summary of the study's methodology and findings. Furthermore, a retrospective analysis of the qualitative study's methodology is conducted, taking into account any constraints and possible interviewer and participant biases. The prospects for more research in this area are then examined.

\subsection{Summary}
The main goal of this study is to provide an overview of the current practices of companies regarding security in software development, to identify trends and patterns, and to position these practices in the context of the literature. For this, three aspects of the software development life cycle are initially identified, guiding the investigation. These aspects are the capture of security requirements, the detection and remediation of security vulnerabilities, and the use of standards.

Existing systematic literature reviews are used to define primary techniques for these elements \ref{chap:Theoretical}. A variety of formal methods are taken into consideration for the purpose of capturing security requirements. Among these, SQUARE \cite{square}, SREP \cite{MELLADO2007244}, MTCSLDC \cite{MTCSDLC} , and CLASP \cite{DEWIN20091152} are the most pertinent. There are recognized key actions in these processes that are used in all processes. These consist of requirements review, risk analysis, and threat modeling. Threat Modeling looks at the software from the standpoint of an attacker in order to find every potential point of failure \cite{XIONG201953}. In Risk Analysis, identified weaknesses and dangers are examined and evaluated based on aspects of the DREAD model, determining the criticality of a vulnerability \cite{1492335} \cite{suprihanto2018determination}. In Requirements Review, a catalog of existing security requirements is filtered based on the properties and requirements of the software being developed, identifying vulnerabilities in the software \cite{MELLADO2007244}.

In handling security vulnerabilities, both detection and subsequent remediation are important. Four primary methods for detecting vulnerabilities are mentioned \cite{8409917}. These are Vulnerability Scanners, where the software is examined for known vulnerabilities listed in vulnerability databases through external scripts \cite{9363884} \cite{8688018}. In Penetration Testing, the software is examined for potential weaknesses through a simulated attack \cite{1392709} \cite{4402456}. In Fuzz Testing, random inputs are passed to the software with the goal of finding unintended behavior, such as crashes or buffer overflows \cite{Klees.2018}. In Static Analysis, the source code is examined for known vulnerabilities, similar to Vulnerability Scanners  \cite{gomes2009overview} \cite{bardas2010static}.

For closing vulnerabilities, two relevant elements are identified. The Incident Response Process describes that a specific procedure should be established, through which the criticality of an emerging vulnerability is quickly determined, necessary steps for remediation are initiated, and finally, an investigation of the approach is conducted for future process improvement \cite{torres2014incident} \cite{AD1180041}. The concept of Emergency Code Response entails the company ensuring that a patch for the problem is provided as quickly as possible  \cite{DISSANAYAKE2022106771}.

Particularly relevant standards identified are ISO 27001, the OWASP guidelines, and the Common Criteria / ISO 15408 \cite{9322704}.

Subsequently, a questionnaire for qualitative, semi-structured interviews is created [\ref{sec:InterviewDesign}]. This is based on the three aspects of the SDLC but also includes additional questions about the company and its attitude towards security to gain better context for the answers. 21 interviews with companies from Germany are conducted, of which 20 are used for evaluation [\ref{sec:InterviewImplementation}].

In the evaluation, the qualitative analysis according to Mayring is followed \cite{mayring1994qualitative}. Various categories are identified in the data, and appropriate coding of the data within these categories is developed [\ref{cap_Data}]. The evaluation of the data reveals that companies give high priority to security in development. It is also recognized that the industry in which the company operates influences the chosen process. Thus, more heavily regulated areas have more and more formal procedures than other companies [\ref{c_Dat_Struc}]. Comparing with the results of the literature, we find that they largely coincide, although the literature sees the processes as more formalized than they are implemented in practice [\ref{c_explication}].

In the last part of the survey, the companies were asked to describe challenges they see in the area of security. It emerged that the primary challenge companies face is the high complexity and rapid development of security, which leads, among other things, to a shortage of skilled personnel in the field.
\subsection{Limitations}
When doing a qualitative study, it is impossible to avoid having both constraints and biases. In the next paragraphs, we would like to discuss the following topics:
\subsubsection{Selection of Participants}
There is a correlation between the interview partners who are chosen and the findings of the study. Because the research largely focused on contacting businesses that already had a connection with the university's department, this implies that there was already a pre-selection of participants, which may have caused the results to be biased. In addition, the size of the sample group places constraints on the level of statistical significance.
\subsubsection{Interviewer Bias} 
It's possible that when drafting the questions and conducting the interviews, certain formulations will be utilized that aim for a particular answer. This is known as interviewer bias. These questions come as a result of the fact that the literature research has provided some probable answers and expectations. Because of this, the replies of the participants may be subconsciously influenced in a particular direction. In a similar vein, this preconceived notion has the potential to result in an incorrect interpretation of the data during the process of evaluating and inspecting the data.
\subsubsection{Participant Bias}
A similar bias can be found among the people who took part in the study. Participants may not make remarks that are entirely truthful regarding security because they do not want to give an unfavorable impression of either themselves or the company. This is because security is a sensitive topic. On the other hand, businesses that regarded their responses as being very unfavorable may have chosen not to take part in the research from the very beginning. In addition, given the delicate nature of the subject matter, there is a possibility that certain questions cannot be answered in their entirety.
\subsection{Future Work}
The objective of this paper is to offer a comprehensive analysis of the situation as it stands right now. This paves the way for further research in a number of different directions. The paper looks at different stages of the software development life cycle (SDLC), although it does not always precisely reflect the particular tactics that companies take during individual periods of development. In subsequent research, this topic may receive a more in-depth examination. One illustration of this would be the methodology behind requirements analysis. An example for this is that the implicit requirements analysis carried out by the companies does not place a primary emphasis on the specific methodology, as was previously explained in section \ref{c_Dat_Struc}. A more in-depth understanding of this stage can be attained through the completion of additional research that focuses on it. Questions that could be asked in response to this include `Which catalogs of security requirements are utilized as a basis?' and `How is the catalog reduced and tailored to the aspects relevant to the software?'

One further possibility is to carry out the study once more,with different companies. This can involve diversifying the industries in order to better recognize possible trends. It is also possible to conduct this study in an international environment to check if this results in different answers.

Lastly, the data collected only allows a vague interpretation as to whether and to what extent ``security by design'' is pursued in today's software development projects. Hence, a follow up study could be designed that focuses on giving answers to this question, for instance by explicitly asking participants to assess which activities are generally carried out in their teams or projects to ensure a high level of security at an early stage in the SDLC.

\bibliographystyle{unsrt}  
\bibliography{paper}

\cleardoublepage
\appendix

\begin{table}[]
    \begin{tabularx}{\textwidth}{|X|r|}
        \hline
        \rowcolor[HTML]{EFEFEF} 
        \textbf{Category}                              & \textbf{Abbreviation}    \\ \hline
        Participant's Role
          & r             \\ \hline
        Company Size                                   & cs       \\ \hline
        Size of the Participant's Development Team 
                          & ts       \\ \hline
        Company's Experience in Software Development
   & ex       \\ \hline
        Company's Target Market
        & tm         \\ \hline
        Types of Software developed by the Company
        & typ         \\ \hline
        Development Process Model followed
        & prc          \\ \hline
        Importance of Security
        & i       \\ \hline
        Development of the Importance of Security
        & ie  \\ \hline
        Means of Security Requirements Elicitation
      & srm    \\ \hline
        SDLC phase of Security Requirements Elicitation
       & srt     \\ \hline
        Responding to Changes in Security Requirements
 & rtc \\ \hline
        Use of  Security Standards
             & std       \\ \hline
        Use of Internal Security Policies
              & pol      \\ \hline
        Use of External Software 
                       & esw          \\ \hline
        Security Assessment Methods
                    & asm      \\ \hline
        Measures in Incident Handling
                & hnd    \\ \hline
        Current and Upcoming Challenges
                & chl       \\ \hline
        Estimated Level of Preparation of the Team or Company to meet the Challenges
                        & prp        \\ \hline
    \end{tabularx}
    \caption{The Categories and their abbreviations we extracted from our data analysis}
    \label{tab:categories}
\end{table}

\begin{table}[th]
    \begin{tabularx}{\textwidth}{|l|X|X|X|X|l|l|l|}
        \hline
        \rowcolor[HTML]{EFEFEF} 
        \textbf{Code} & \textbf{r} & \textbf{cs} & \textbf{ts} & \textbf{ex} & \textbf{tm}     & \textbf{typ} & \textbf{prc} \\ \hline
        \textbf{A}    & Developer     & \textless 500       & \textless 8         & \textless 14        & Finance               & Web               & Agile            \\ \hline
        \textbf{B}    & Architect     & 500 - 1000          & 8 - 10              & 14 - 25             & Online Commerce       & Infrastructure    & Waterfall        \\ \hline
        \textbf{C}    & Lead          & 1001 - 5000         & 11 - 12             & 26 - 30             & Public Infrastructure & Control Devices   & V-Model          \\ \hline
        \textbf{D}    & -             & \textgreater 5000   & \textgreater 12     & \textgreater 30     & Marketing             & CMS               & -                \\ \hline
        \textbf{E}    & -             & -                   & -                   & -                   & Automotive            & Mobile Apps       & -                \\ \hline
        \textbf{F}    & -             & -                   & -                   & -                   & Industrial Software   & Mixed             & -                \\ \hline
        \textbf{G}    & -             & -                   & -                   & -                   & Mixed                 & Unclear           & -                \\ \hline
        \textbf{H}    & -                   & -                              & -                             & -                             & -                             & -                             & - \\ \hline
        \textbf{I}    & -                   & -                              & -                             & -                             & -                             & -                             & - \\ \hline
        \textbf{J}    & -                   & -                              & -                             & -                             & -                             & -                             & - \\ \hline
    \end{tabularx}
    \caption{First set of codes}
    \label{tab:codes_1}
\end{table}

\begin{table}[th]
    \begin{tabularx}{\textwidth}{|l|X|X|X|X|X|}
        \hline
        \rowcolor[HTML]{EFEFEF} 
        \textbf{Code} & \textbf{i}         & \textbf{ie} & \textbf{srm} & \textbf{srt} & \textbf{rtc} \\ \hline
        \textbf{A}    & Highest                     & Constant                 & Dedicated Analysis          & Analysis              & Go back to Analysis       \\ \hline
        \textbf{B}    & Equivalent to functionality & Increasing               & Security Standards          & Design                & Agile                     \\ \hline
        \textbf{C}    & Less than functionality     & -                        & Intuitive                   & Agile                 & -                         \\ \hline
        \textbf{D}    & -                           & -                        & Raised by Clients           & -                     & -                         \\ \hline
        \textbf{E}    & -                   & -                              & -                             & -                             & -                 \\ \hline
        \textbf{F}    & -                   & -                              & -                             & -                             & -                 \\ \hline
        \textbf{G}    & -                   & -                              & -                             & -                             & -                 \\ \hline
        \textbf{H}    & -                   & -                              & -                             & -                             & -                 \\ \hline
        \textbf{I}    & -                   & -                              & -                             & -                             & -                 \\ \hline
        \textbf{J}    & -                   & -                              & -                             & -                             & -                 \\ \hline
    \end{tabularx}
    \caption{Second set of codes}
    \label{tab:codes_2}
\end{table}

\begin{table}[th]
    \begin{tabularx}{\textwidth}{|l|X|X|X|X|}
        \hline
        \rowcolor[HTML]{EFEFEF} 
        \textbf{Code} & \textbf{std} & \textbf{pol}           & \textbf{esw}              & \textbf{asm} \\ \hline
        \textbf{A}    & ISO 27001           & With Coding Guidelines    & Whitelist                     & Code Analysis            \\ \hline
        \textbf{B}    & OWASP               & Without Coding Guidelines & Supported Software            & CVE Scanning             \\ \hline
        \textbf{C}    & BSI                 & -                              & Explicit Security Check       & Penetration Testing      \\ \hline
        \textbf{D}    & Industry-specific   & -                              & No or informal Security Check & Compliance Testing       \\ \hline
        \textbf{E}    & Defined by Client   & -                              & -                             & Threat Modeling          \\ \hline
        \textbf{F}    & -                   & -                              & -                             & External Audits          \\ \hline
        \textbf{G}    & -                   & -                              & -                             & SBOM                     \\ \hline
        \textbf{H}    & -                   & -                              & -                             & -                        \\ \hline
        \textbf{I}    & -                   & -                              & -                             & -                        \\ \hline
        \textbf{J}    & -                   & -                              & -                             & -                        \\ \hline
    \end{tabularx}
    \caption{Third set of codes}
    \label{tab:codes_3}
\end{table}

\begin{table}[t]
    \begin{tabularx}{\textwidth}{|l|X|X|X|}
        \hline
        \rowcolor[HTML]{EFEFEF} 
        \textbf{Code} & \textbf{hnd}       & \textbf{chl}        & \textbf{prp}     \\ \hline
        \textbf{A}    & Investigate Criticality           & High Complexity            & Well prepared        \\ \hline
        \textbf{B}    & Develop and Release Hotfix        & Rapid Development          & Fairly well prepared \\ \hline
        \textbf{C}    & Deactivate Features in Software   & High Costs                 & Somewhat prepared    \\ \hline
        \textbf{D}    & Check other Projects for Incident & Requires Expertise         & Not well prepared    \\ \hline
        \textbf{E}    & Automated Dependency Updates      & Lack of Awareness          & -                    \\ \hline
        \textbf{F}    & Inform Security Authorities       & Hard to be Compliant       & -                    \\ \hline
        \textbf{G}    & Conduct Feedback Process          & Deciding on right Tools    & -                    \\ \hline
        \textbf{H}    & -                                 & Constantly new Threats     & -                    \\ \hline
        \textbf{I}    & -                                 & Conflicting Priorities     & -                    \\ \hline
        \textbf{J}    & -                                 & Missing Design Perspective & -                    \\ \hline
    \end{tabularx}
    \caption{Fourth set of codes}
    \label{tab:codes_4}
\end{table}

\begin{sidewaystable}
	\begin{tabular}{|l|l|l|l|l|l|l|l|l|l|l|l|l|l|l|l|l|l|l|l|}
\hline
\rowcolor[HTML]{EFEFEF} 
\textbf{ID} & \textbf{r} & \textbf{cs} & \textbf{ts} & \textbf{ex} & \textbf{tm} & \textbf{typ} & \textbf{prc} & \textbf{i} & \textbf{ie} & \textbf{srm} & \textbf{srt} & \textbf{rtc} & \textbf{std} & \textbf{pol} & \textbf{esw} & \textbf{asm}    & \textbf{hnd} & \textbf{chl} & \textbf{prp} \\ \hline
1                      & B             & C                & -                & A            & A            & A             & A; B          & A            & A            & B              & A              & B            & B; C; D       & B            & A; B; C        & D; F; I; J       & A; B; C; G    & A; C          & A             \\ \hline
2                      & B             & A                & -                & A            & B            & B             & A             & B            & A            & A              & A              & A            & A             & A            & D              & A; B; D; J       & A; B; C; F; G & A; B          & C             \\ \hline
3                      & C             & B                & -                & D            & C            & A; C          & A; B          & B            & B            & B              & B              & -            & A; B          & B            & C              & A; B; C; G; I; J & A; B          & A             & B             \\ \hline
4                      & A             & D                & B                & B            & D            & D             & A             & B            & B            & A              & B              & A            & A; E          & B            & D              & A                & B             & E; G          & C             \\ \hline
5                      & C             & D                & -                & -            & E            & A; C          & A             & A            & A            & A              & A              & B            & A; D          & B            & C              & B; C; E; F; I; J & A; B; C; D; G & E; F          & B             \\ \hline
6                      & A             & B                & C                & D            & C            & A; C          & A             & A            & B            & B              & B              & B            & A; D          & B            & B; D           & B; D; J          & A             & A; D; E       & B             \\ \hline
7                      & A             & B                & B                & D            & C            & A; C          & A             & B            & B            & B              & B              & B            & -             & B            & -              & B; C; D          & A; G          & A; D; E       & B             \\ \hline
8                      & B             & B                & B                & C            & G            & A             & A             & C            & B            & C              & C              & B            & E             & B            & -              & A                & A; B; E       & A; C; G       & D             \\ \hline
9                      & C             & C                & C                & C            & F            & E             & A; B          & B            & B            & -              & -              & A            & -             & B            & A; C           & F                & A; B; G       & D             & C             \\ \hline
10                     & B             & D                & D                & -            & E            & A; C          & A             & A            & B            & A              & A              & B            & D             & B            & C              & B; C; F; I; J    & A; B; C; D; G & F             & B             \\ \hline
11                     & C             & A                & C                & B            & E            & F             & A             & C            & -            & C              & C              & B            & B             & B            & D              & A; B; C; E; I; J & D; G          & E; G          & B             \\ \hline
12                     & B             & D                & D                & A            & G            & G             & A             & B            & A            & C              & C              & B            & A             & B            & D              & A; B; D; F; I    & A; B; D       & E; F          & C             \\ \hline
13                     & A             & C                & D                & A            & A            & A             & A             & A            & B            & B              & C              & B            & B; D          & A            & D              & A; B; C; I; J    & A; B          & A; B; G       & B             \\ \hline
14                     & C             & A                & A                & B            & G            & A             & A             & B            & B            & A; D           & A              & B            & A             & A            & D              & A; B; C; E; I; J & A; B; E       & B; F; H       & C             \\ \hline
15                     & B             & A                & C                & B            & E            & F             & A             & B            & B            & A              & A              & B            & A             & -            & -              & A; C; D; E       & -             & I             & B             \\ \hline
16                     & C             & D                & -                & C            & G            & F             & A             & B            & B            & D              & A              & B            & A; D          & A            & C              & A; B; C; I; J    & A; B; F; G    & E; F          & B             \\ \hline
17                     & A             & A                & A                & A            & G            & A             & A             & B            & B            & C              & C              & B            & A             & B            & D              & A; C; I          & B; D; E       & D             & B             \\ \hline
18                     & B             & B                & B                & C            & C            & A             & A             & C            & B            & C              & C              & B            & A             & B            & D              & A; B; C; G       & B; D          & D; E; J       & D             \\ \hline
19                     & C             & B                & B                & C            & C            & A             & A             & B            & B            & A              & A              & B            & A             & B            & D              & A; B; C; I       & A; B; G       & B; F; H       & D             \\ \hline
20                     & B             & C                & -                & C            & A            & A             & A; C          & A            & A            & B              & A              & A            & C             & -            & D              & A; B; C; E; F    & -             & E             & -             \\ \hline
	\end{tabular}
     \caption{Distribution of codes from the interviews.}
    \label{tab:codes_5}
\end{sidewaystable}

\end{document}